\documentclass[a4paper]{article} 
\usepackage{cite}
\usepackage{authblk}
\usepackage{amsmath,amssymb,amsfonts,amsthm}
\usepackage{algorithmic}
\usepackage{graphicx}
\usepackage{tabularx}
\usepackage{multirow}
\usepackage{textcomp}
\usepackage[dvipsnames,table]{xcolor}
\usepackage{array}
\usepackage{url}
\usepackage{mdframed}
\usepackage[most]{tcolorbox}
\usepackage{hyperref}
\usepackage[colorinlistoftodos]{todonotes}
\usepackage[margin=1in]{geometry}
\usepackage{enumitem}
\usepackage{hhline}
\usepackage[T1]{fontenc}
\usepackage{array}
\usepackage{calc}
\newcolumntype{C}{>{\centering\arraybackslash}X}

\newtheoremstyle{boldstyle}
  {\topsep}
  {\topsep}
  {\itshape}
  { }
  {\bfseries}
  {.}
  { }
  {\thmname{#1}\thmnumber{ #2}\thmnote{ (#3)}}
\theoremstyle{boldstyle}
\newtheorem{openquestionx}{Open Question}

\newmdenv[
  skipabove=2pt,
  skipbelow=2pt,
  linewidth=1pt,
  linecolor=blue,
  backgroundcolor=blue!5,
  roundcorner=5pt,
  innertopmargin=2pt,
  innerbottommargin=2pt,
  innerleftmargin=3pt,
  innerrightmargin=3pt
]{openboxq}
\newenvironment{openquestion}
  {\begin{openboxq}\begin{openquestionx}}
  {\end{openquestionx}\end{openboxq}}

\newtheorem{definitionx}{Definition} 
\newmdenv[
  skipabove=2pt,
  skipbelow=2pt,
  linewidth=1pt,
  linecolor=orange,
  backgroundcolor=orange!5,
  roundcorner=5pt,
  innertopmargin=2pt,
  innerbottommargin=2pt,
  innerleftmargin=3pt,
  innerrightmargin=3pt
]{defopenboxq}

\newenvironment{definition}
  {\begin{defopenboxq}\begin{definitionx}}
  {\end{definitionx}\end{defopenboxq}}

\def\BibTeX{{\rm B\kern-.05em{\sc i\kern-.025em b}\kern-.08em
    T\kern-.1667em\lower.7ex\hbox{E}\kern-.125emX}}
    
\newcommand{\cm}[1]{\textcolor{blue}{\textbf{Conor:} #1}}
\newcommand{\qb}[1]{\textcolor{red}{\textbf{Quentin:} #1}}

\newcommand{\dk}[1]{\textcolor{cyan}{\textbf{Demetris:} #1}}

\newcommand{\open}[1]{\textcolor{green}{\textbf{open:} #1}}

\title{SoK: Preconfirmations\footnote{© 2026 IEEE.  Personal use of this material is permitted.  Permission from IEEE must be obtained for all other uses, in any current or future media, including reprinting/republishing this material for advertising or promotional purposes, creating new collective works, for resale or redistribution to servers or lists, or reuse of any copyrighted component of this work in other works.}}

\author[1]{Aikaterini-Panagiota Stouka}
\author[1]{Conor McMenamin}
\author[3]{Demetris Kyriacou}
\author[1]{Lin Oshitani}
\author[2]{Quentin Botha}

\affil[1]{\small\textit{Nethermind Research}}
\affil[2]{\small\textit{Research Institute for Cryptoeconomics, WU Vienna}}
\affil[3]{\small\textit{Imperial College London}}

\begin{document}

\maketitle

\begin{abstract}
 
In recent years, significant research efforts have focused on improving blockchain throughput and confirmation speeds without compromising security. While decreasing the time it takes for a transaction to be included in the blockchain ledger enhances user experience, a fundamental delay still remains between when a transaction is issued by a user and when its inclusion is confirmed in the blockchain ledger. This delay limits user experience gains through the confirmation uncertainty it brings for users. This inherent delay in conventional blockchain protocols has led to the emergence of preconfirmation protocols -- protocols that provide users with early guarantees of eventual transaction confirmation.

This article presents a Systematization of Knowledge (SoK) on preconfirmations\footnote{Another version of this document is hosted on GitHub at: https://github.com/NethermindEth/sok-preconfirmations. This paper has been accepted to IEEE ICBC 2026. The accepted version TBA.}. We present the core terms and definitions needed to understand preconfirmations, outline a general framework for preconfirmation protocols, and explore the economics and risks of preconfirmations. Finally, we survey and apply our framework to several implementations of real-world preconfirmation protocols, bridging the gap between theory and practice.

\end{abstract}

\setlength{\marginparwidth}{2cm} 

\section{Introduction}

In recent years, the scalability of blockchain systems has been driven by advances within blockchain protocol design (see \cite{cryptoeprint:2018/1119, che2025manifoldchain, Fitzi2020ProofofStakeBP, neiheiser2025anthemiusefficientmodular, ethereum_scaling}), but also by out-of-protocol innovations in the transaction lifecycle.
A particularly important yet relatively underexplored stage of this lifecycle is the period between when a transaction is submitted through a wallet and when it is confirmed on-chain. During this interval, users face uncertainty about if, when, and/or how their transactions will appear on the blockchain. This window of uncertainty has motivated the emergence of preconfirmation protocols, through which users can receive stronger assurances about transaction inclusion or ordering before final consensus is reached.

Preconfirmations have rapidly gained traction across blockchains, particularly on additional layers built on top of them such as Layer 2s (see \hyperref[sec:background]{Section \ref{sec:background}}). There, preconfirmations serve diverse goals ranging from improving user experience to enabling new application primitives \cite{Optimism,Arbitrum,ZKsync, W:TaikoPreconfDesign, W:JitoSolanaShreds}. For example, traders desire predictable execution in DeFi, wallets seek to minimize friction for mainstream adoption, and cross-domain systems depend on timely bridging guarantees. Despite the growing relevance of preconfirmations, the design space remains fragmented, with projects re-inventing solutions in isolation and often overlooking subtle trade-offs between trust, latency, and decentralization.

This article presents a Systematization of Knowledge (SoK) on preconfirmations. We survey the landscape of existing approaches, categorize their underlying mechanisms, and highlight both the promises and pitfalls of this rapidly evolving space. By placing current designs within a broader taxonomy, we aim to clarify their assumptions, identify unaddressed challenges, and provide a framework for reasoning about future directions.

Ultimately, preconfirmations should be seen as a structural component of blockchain systems with implications for protocol security, economics, and blockchain adoption. This work contributes a unified lens for evaluating the role of preconfirmations, setting the stage for deeper inquiry into how preconfirmations can be integrated in creating predictable, secure, and user-friendly transaction pipelines.

\subsection{Organization of the SoK}
The structure of the main body of the SoK is as follows:
\begin{itemize}
    \item In \hyperref[sec:background]{Section \ref{sec:background}}, we present the fundamental concepts of blockchain that are essential for understanding preconfirmation protocols.
    \item In \hyperref[sec:definitions]{Section \ref{sec:definitions}}, we present the fundamental definitions and terms used in preconfirmation protocols (e.g., what a preconfirmation is, its types, and who the entities in a preconfirmation protocol are). 
    \item In \hyperref[sec:pipeline]{Section \ref{sec:pipeline}}, we outline the structure of a preconfirmation protocol as a sequence of steps and describe how different types of preconfirmation protocols vary at each step.
    \item In \hyperref[preconf_economics]{Section \ref{preconf_economics}}, we focus on research that has been/needs to be conducted on the economics of preconfirmation protocols. This includes, for example, studies on the revenues of entities that provide guarantees to users, as well as analyses of how users should reward these entities with tips to incentivize their participation in the protocol and encourage honest behavior.
    \item In \hyperref[sec:risk]{Section \ref{sec:risk}}, we discuss the risks that can arise from preconfirmation protocols -- both for the entities involved in the protocol (e.g., users requesting guarantees and entities providing them) and for the underlying blockchain protocol itself.
    \item Equipped with a general understanding of preconfirmation protocol concepts, in \hyperref[sec:implementations]{Section \ref{sec:implementations}} we present several preconfirmation protocols that are currently in production. When a protocol differs slightly from the skeleton described in \hyperref[sec:pipeline]{Section \ref{sec:pipeline}}, we explain how it deviates from the general structure.
    \item We conclude in \hyperref[sec:conclusion]{Section \ref{sec:conclusion}}.
    \item In \hyperref[Appendix_A]{Appendix \ref{Appendix_A}}, we discuss how preconfirmations can be formulated and provided for intents, as opposed to transaction-based preconfs.
\end{itemize}

\section{Related Work}

This SoK builds on previous work conducted to aggregate and organise preconfirmation knowledge. Most notably, \cite{W:AwesomeBasedPreconfirmations} contains a collection of resources and articles related to preconfirmations that are periodically updated. Attempts have also been made to explain fundamental preconfirmation concepts \cite{W:PreconfirmationsGlossaryRequirements, W:Preconfirmations:Explained}. This SoK builds on these works, standing as a complete guide to all preconfirmation-related concepts.

\section{Background} 
\label{sec:background}
This section outlines the established blockchain concepts necessary to understand preconfirmations. Although preconfirmations are blockchain-agnostic, in this document, we restrict our discussion to preconfirmations on Ethereum.

    \subsection{Blockchain Fundamentals}
    \label{subsec:BlockchainFundamentals}
    For the purpose of this document, we consider a \textbf{blockchain} as both the protocol and data structure used to derive a replicated state machine among a set of distributed, potentially distrusting, peers/nodes. Although the blockchain protocol defines rules 
    intended to produce a single shared state for observers who follow the blockchain protocol, conflicting states can still emerge temporarily. \textbf{Consensus mechanisms} are used to resolve these conflicts and determine the single shared state of the system. In this document, we assume blockchains derive a single shared state. Furthermore, we focus on transaction-based blockchains which form the majority of blockchains today. These blockchains advance the shared state by sequentially executing an ordered list of transactions organized in a block. Every block links to the previous block. The canonical list of transactions that derive the single shared state is denoted as the \textbf{blockchain ledger}. \textbf{Inclusion} is the process of adding a transaction to the ordered list of transactions, and \textbf{execution} is the computation of a transaction's result, thereby updating the shared state. This distinction is critical for delineating the types of preconfirmations that can be given.

    Many of the most popular blockchains fall under the category Proof of Stake (PoS) blockchains. In PoS blockchains, the ledger is extended by entities known as validators, who hold stake in the blockchain -- that is, they possess some amount of the blockchain’s native currency. These validators are typically selected to extend the ledger with a probability proportional to the size of their stake. 
    
    In Ethereum, which belongs to this category of blockchains, validators need to lock a minimum amount of the native currency, called Ether (ETH), to be able to participate in extending the blockchain ledger. This locked amount is referred to as \textbf{collateral} or deposit. In addition to proposing new blocks, validators also attest to (vote on) the blocks proposed by other validators. Validators attesting to or proposing a block receive rewards minted by the system. In addition, they can earn rewards through the tips of the transactions they include in the block they propose. These tips are known as \textbf{transaction fees}. If validators behave maliciously in verifiable ways specified by the rules of the protocol, they get penalized by losing a portion of their collateral -- a process known as \textbf{slashing}. 
    
    Ethereum is a smart-contract enabled blockchain. \textbf{Smart contracts} can be considered as a collection of code/functions, data, and state within the global state of a blockchain~\cite{W:IntroductionToSmartContracts} that enable applications to be built on and trustlessly executed by transactions on the underlying blockchain. At a high level, a smart contract consists of an algorithm and public data, both of which are created and updated through transactions originated by users and executed by validators. In the context of preconfirmations, smart contracts serve as critical components for coordinating and enforcing preconfirmations.

    \subsection{Ethereum's Block-Building Pipeline}
    \label{sec:L1_pipeline}
    
    In \hyperref[sec:L1_pipeline]{Section~\ref{sec:L1_pipeline}}, we highlight the key components of the Ethereum blockchain that are critical for understanding preconfirmations. In more detail, we describe the process through which blocks are built and eventually proposed, known as the block-building pipeline. 
    
    \paragraph{Transaction Propagation and Mempools}
        Users create transactions through their digital wallet applications. These transactions are then sent to nodes, which maintain transaction mempools and expose standardized interfaces called Remote Procedure Call (RPC) endpoints. The nodes perform basic validation checks on each transaction (e.g., verifying the sender’s balance) before adding it to their mempool and potentially propagating it across the peer-to-peer (P2P) network. Nodes that do not propagate certain transactions are said to retain a private mempool. Validators select transactions from the mempool to construct an ordered list of transactions for inclusion in their blocks. Even if a transaction passes the initial validation checks before entering the mempool, it may still become invalid during execution if, for example, a prior transaction in the mempool drains the sender’s balance.  
        
    \paragraph{Proposers and Attesters: Selection and Responsibilities} \label{proposersAndAttesters}
Ethereum progresses in epochs each consisting of 32 slots of $12$ seconds. During every slot, there is one eligible validator selected to construct, sign, and propose a block. This validator, known as proposer, is pseudo-randomly selected from the set of registered validators. Beyond the rewards earned by following the Ethereum protocol (see \hyperref[subsec:BlockchainFundamentals]{Section \ref{subsec:BlockchainFundamentals}}), the proposer can extract additional value -- known as maximal extractable value (MEV)~\cite{W:MaximalExtractableValueMEV} -- by intentionally manipulating the list of transactions proposed within the slot. For example, the trade prices offered by a type of decentralized exchange called Automated Market Maker (AMM) change after every trade. A proposer can profit from a user's buy order by including their own buy order in front of the user's buy order -- \textbf{front-running} the buy order -- and/or including their own sell order behind the user's buy order -- \textbf{back-running} the user's buy order. Together, front-running and back-running the same transaction is known as \textbf{sandwiching}~\cite{W:MaximalExtractableValueMEV}. Apart from proposing blocks, validators are elected into committees that vote on the validity (as specified by the Ethereum protocol) of proposed blocks -- a process known as attesting. 

The task of constructing a block can be unbundled from the tasks of signing and proposing the block. This is known as Proposer-Builder Separation (PBS) \cite{W:Proposer-builderseparation}. PBS enables proposers to delegate the construction of their block to specialized entities known as builders.  
In PBS, proposers select the transaction list to be proposed, referred to as the \texttt{ExecutionPayload} (see \cite{W:TheMerge--TheBeaconChain}), by running an auction among builders. Builders bid in the auction, competing to have their own transaction lists proposed, with the winning payment paid to the proposer. Note that when the proposer selects a transaction list, they can only see a commitment to it, not the entire list. The complete list is revealed only after the proposer has made their selection. Another entity, known as the relayer (not to be confused with preconfirmation gateways introduced later), provides fair exchange \cite{P:Fairexchangewithasemi-trustedthirdparty} of the committed transaction list between proposers and builders. MEV-Boost \cite{MEV-Boost}, an implementation of PBS for Ethereum, is the primary source of blocks on Ethereum with over 90\% of blocks being sourced through MEV-Boost at the time of writing \cite{MEV.pics}. 

    \paragraph{Proposer Lookahead} \label{proposerLookahead}
        The proposer lookahead specified by the Ethereum protocol provides advance notice to validators of their proposing duties.
        According to EIP-7917 \cite{EIP7917} which is a candidate for inclusion in Ethereum’s upcoming Fusaka upgrade~\cite{W:EIP-7607:HardforkMeta-Fusaka} the proposer schedule for epoch~$N+1$ becomes known and accessible -- e.g., via smart contracts -- at the start of epoch~$N$. 
        This lookahead is critical for enabling preconfirmations, as it allows future proposers to begin offering preconfirmations before their assigned slot arrives. \\

\subsection{Ethereum Scaling Solutions}\label{sec:intro_L2}
To enable a decentralized network to keep up with and reach consensus on state updates in blockchains, throughput is more limited compared to centralized solutions where a single trusted party maintains and updates transaction records. To increase throughput while still preserving key features of the blockchain (e.g., decentralization), scaling solutions in the form of additional layers have emerged. The original blockchain, in our case Ethereum, will be denoted by layer 1 (L1), and its canonical list of blocks with transactions by \textbf{L1 blockchain ledger}. The other layers will be called layer 2 (L2s), L3s, or generally LNs for some $N>1$. These LNs are execution environments that typically move one or more of the L(N-1) resources outside of the critical path of L(N-1) state progression. At the time of writing, the majority of LN users exist at L2, so without loss of generality, we will focus on L2s when discussing these blockchain scaling solutions. To the best of our knowledge, all statements about L2s and their relationship to the L1 can apply to any LN and its respective L(N-1). 
\par
Deploying as an L2 instead of L1 removes the need for the L2 to establish its own consensus protocol or dedicated economic security for securing state transitions (although these features can be added). L2s can tap into the user base of the L1, allowing L1 users to opt in to locking their funds and using them within the L2. Many L2 solutions exist, including rollups, validiums, and Plasmas \cite{L2_versus_execution_sharding}. 
As in L1, L2 state progresses through proposers proposing blocks, establishing a canonical list of blocks, and some eventual finalization of these blocks/commitments to these blocks on L1 to form the \textbf{L2 blockchain ledger}. L2 state is then derived from this L2 blockchain ledger. L2 solutions can differ in how they finalize L2 state or in the type of data they post to L1. However, these details are beyond the scope of this document. With respect to preconfirmations, two key L2-specific concepts are relevant:
\begin{enumerate}
    \item \textbf{L2 proposer selection:} A key factor impacting preconfs is whether an L2 is based or non-based.
    \begin{enumerate}
        \item \textbf{Based L2:} Delegates block proposing to the proposers of L1 ~\cite{W:ExaminingtheBasedSequencingSpectrum,W:BasedrollupssuperpowersfromL1sequencing}. 
        \item \textbf{Non-based L2:} Uses a dedicated proposer election mechanism. Within this category of non-based L2s, a key differentiating factor is whether there is one or many proposers. When there is only one proposer, preconfirmations are greatly simplified, at the cost of centralization, and the negative effects that this brings (e.g., single point of failure, monopolization). These single proposer non-based L2s are, rather appropriately, sometimes referred to as centralized sequencer L2s.  Some examples of non-based L2s include Optimism \cite{Optimism}, Arbitrum \cite{Arbitrum}, ZKsync \cite{ZKsync}, and Starknet \cite{Starknet}.
    \end{enumerate}
    \item \textbf{L2 governance:}
    Governance is a protocol for determining how a set of designated entities can modify some or all of the blockchain protocol specification. These permissions may stem from being a founder, a stakeholder, or being elected within the protocol. Governance can have many roles, including upgrading the chain, adjusting protocol parameters, or selecting entities for certain privileged roles -- such as proposer or overseer (introduced in \hyperref[def:overseer]{Definition \ref{def:overseer}}) -- within the blockchain protocol (see~\cite{W:ThestateofArbitrumsprogressivedecentralization,W:Starknetstoken:STRK,W:GovernanceinSeason8:TheNextPhase}).     
    In Ethereum, the rules governing how blockchain ledger progression and state derivation change via social consensus among validators. This governance-related consensus occurs outside of Ethereum's protocol rules. 
\end{enumerate}

\section{General Definitions and Concepts of Preconfirmations}
\label{sec:definitions}
With this background in hand, we are ready to introduce and discuss preconfirmations. At a high level, a preconfirmation in the blockchain context is a promise/commitment that the blockchain ledger will satisfy some property at some point in the future. The overwhelming majority of preconfirmations being considered, both in production and development, involve promises about the inclusion of a specific transaction or transactions in the transaction list of a block that will eventually be part of the blockchain ledger.

\par To formally define preconfirmations, we use the concept of logical predicates. A predicate function determines whether a given input to the function possesses a specific property. In the blockchain context, a predicate function taking as input a blockchain ledger, as formed at a specific point in time, e.g., a slot in Ethereum, would return true if and only if the property defined by the predicate function is satisfied by the blockchain ledger. In this context, we define a preconfirmation as follows:
 


        \begin{definition}
        \label{def:preconfer}
        For a given blockchain, a \textbf{preconfirmation (preconf)} is a commitment to some predicate function $\mathsf{f}$ such that: there exists a point in time -- after the preconfirmation is issued -- such that if the blockchain ledger at that time is given as input to the predicate, it will return true. 
        \end{definition}

        For preconfs, the commitment can be considered as a promise. The exact structure, significance, and strength of the commitment depends on the individual protocol, and entity providing the commitment.
        For example, a commitment from an upcoming proposer is different to a commitment from an entity who does not control proposal rights -- the proposer can directly influence the blockchain ledger. A commitment from an entity who is financially punished for not satisfying the commitment is different to a commitment from an entity who faces no repercussion for not satisfying the commitment. Who can offer preconfs, the relative strength of preconf commitments, preconf incentives, and preconf enforcement mechanisms are expanded on throughout the remainder of this document. 

        Additionaly, we define \textbf{conditional preconfs} as preconfs for which the corresponding promise is valid only if some pre-defined condition is met. The condition is also a predicate that takes as input the blockchain ledger.

        \begin{definition}
            A \textbf{conditional preconf} for a promised predicate  $\mathsf{f}$ and a conditional predicate $\mathsf{c}$ is a commitment that if the condition $\mathsf{c}$ is satisfied by the blockchain ledger in the future, then the predicate function $\mathsf{f}$ will also be satisfied by the blockchain ledger at some point after the preconf is issued. 
        \end{definition}
        For example, a conditional preconf could promise that a transaction will be included in the blockchain ledger if a specific block builder is selected to build the block for a given slot. This is the type of preconf builders can offer in mev-commit~\cite{W:Documentation-Understandingmev-commit}. Another example of a conditional preconf could be a promise that a transaction will be included in the blockchain ledger if the base fee (i.e, minimum required fee payable to the protocol) in a given block is below some threshold. 

         Preconfs, whether conditional or not, are only meaningful if the committed predicates can be satisfied. To describe this notion of predicate satisfaction, we introduce the concept of fulfillment, which holds for all preconfs, including conditional preconfs, as follows: 

     \begin{definition}\label{def:delivery}
    A preconf is considered \textbf{fulfilled }if and only if the promised predicate specified by the preconf is satisfied by the blockchain ledger. Satisfying a preconf's promised predicate is equivalent to fulfilling the preconf.
    \end{definition}


        Preconfs can be implemented in many different ways. Key factors that distinguish preconf implementations include who is providing the preconf, the predicate involved in the preconf, and blockchain governance, if it exists. The type of preconf determines which applications and users benefit from them.  We distinguish and explain these varying preconf implementations throughout this document. In this document, we will focus on transaction-based preconfs, defined in the following section. 
        
        Other types of preconfs not requiring commitments to specific transactions or transaction sequences are possible. These are generally referred to as intent preconfs. An example of an intent is a request to exchange a specific amount of token A for some minimum amount of token B without a specific user transaction. Although we focus on transaction-based preconfs in the main part of this document, we provide some information on intent preconfs in Appendix~\ref{Appendix_A}.
     
    \subsection{Transaction-based preconfs}
    \label{subsec:transaction_based}
    Transaction-based preconfs are preconfs with predicates that depend on the existence of a transaction or sequence of transactions in the blockchain ledger. Some examples of this type of predicate are the following: 
    \begin{enumerate}
        \item A given transaction is included in the blockchain ledger in an arbitrary position.
        \item A given transaction appears at a specific position in the blockchain ledger, e.g., after a specific sequence of transactions.
    \end{enumerate}
    Below, we formally define different types of preconfs based on the structure of their predicates.

    The main types of preconfs studied in the literature are \textbf{inclusion} and \textbf{execution} preconfs (see \cite{W:PreconfirmationsGlossaryRequirements,W:PreconfirmationsforVanillaBasedRollups,W:APricingModelforInclusionPreconfirmations,W:AnalysingExpectedProposerRevenuefromPreconfirmations}), defined as follows:   
    
        \begin{definition}
        For a given blockchain and transaction $\mathrm{tx_0}$, an \textbf{inclusion preconf} is a preconf where the predicate function returns true if $\mathrm{tx_0}$ is included in the blockchain ledger given as input.
        \end{definition}        
        Generally, inclusion preconfs are useful when the primary concern of the user or application is ensuring that the transaction is included. Use cases include simple transfers or posting sequences of L2 transactions (batches) to L1.  
        However, inclusion preconfs provide no guarantee of how a transaction executes, which becomes problematic for transactions seeking to act on \textbf{contentious state}~\cite{W:AnalysingExpectedProposerRevenuefromPreconfirmations,W:APricingModelforInclusionPreconfirmations}. 
        \begin{definition}
        \label{def:contentious state}
        \textbf{Contentious state} is a state element upon which concurrent transactions attempt to act, creating a race condition among transactions to act on the state element first. This state element may be derived directly from the existing blockchain ledger, or from transactions that are expected to be included in the blockchain ledger (e.g., preconfed transactions building on the blockchain ledger)  \end{definition}
        
        Examples of contentious state include arbitrage and liquidation opportunities on a blockchain. For a transaction trying to act on such opportunities, or contentious state in general, execution preconfs are more comprehensive commitments than inclusion preconfs~\cite{W:ATaxonomyofPreconfirmationGuaranteesandTheirSlashingConditionsinRollups}.
        \begin{definition}
        \label{def:execution_preconf}
        For a given blockchain, transaction $\mathrm{tx_1}$ and sequences of transactions $\mathrm{stx}_1$, an \textbf{execution preconf} is a preconf where the promised predicate is a function that returns true if $\mathrm{tx_1}$ is included in the blockchain ledger immediately after $\mathrm{stx}_1$.
        \end{definition}
        We refer to transactions that have received inclusion preconfs or execution preconfs as inclusion preconfed and execution preconfed transactions, respectively. When the context of inclusion preconf or execution preconf is clear or unimportant, we simplify this terminology to just preconfed transactions.
        
        Providing execution preconfs is more complex and comes with additional block-building constraints compared to providing inclusion preconfs. Execution preconfs can only be issued either (i) for the current slot, or (ii) for a future slot, provided that the transactions to be included in all preceding slots have already been preconfed. This is because execution preconfs require knowledge of the sequence of transactions in the blockchain ledger preceding the transaction to be preconfed.

    \subsection{Preconfers}
    \label{sec:preconfirmers_preliminaries}
        Preconfs are provided by preconfers. The entity that can perform the preconfer role depends on the layer and sequencing model considered. 
        \begin{definition}
        \label{def:preconfirmer}
        A
        \textbf{preconfer} is an entity that provides preconfs, with \textbf{preconfing} being the act of issuing a preconf by a preconfer.  
        \end{definition}
        
        Preconfers have varying abilities to fulfill preconfs. Block proposers -- whether for L1 or L2 -- are candidates for preconfers. This is because proposers can enforce fulfillment of their preconfs by virtue of their control over block construction and proposal.
        Although proposers have the option to act as preconfers -- since they are entities capable of fulfilling preconfs -- doing so may increase their operational complexity.
        Proposers can overcome this complexity by delegating their preconfing duties to gateways.
        \begin{definition}
        \label{def:preconfirmation}
        In the context of preconfs, a \textbf{gateway} is a specialized entity that preconfs on behalf of a proposer.
        \end{definition}
        As gateways lack direct control over block proposal/signing, preconfs sourced from gateways depend on the ability of the gateway to enforce fulfillment by the proposer~\cite{W:ThePreconfirmationGatewayUnlockingPreconfirmations:FromUsertoPreconfer, W:Ahead-of-TimeBlockAuctionsToEnableExecutionPreconfirmations,W:DelegationinBolt:OutsourcingSophisticationWhilePreservingDecentralization}. How exactly gateways fulfill preconfs to the proposer is the focus of \hyperref[preconf_delivery]{Section~\ref{preconf_delivery}}.
        
        Another preconfer candidate is builders in the current MEV-Boost PBS paradigm. Builders as preconfers is one of the core focuses of the mev-commit preconf protocol \cite{W:Documentation-Understandingmev-commit}. Recall that in MEV-Boost, builders construct blocks and bid for the right for their block to be selected for proposal by the proposer. Given a builder must construct a block without knowing whether or not the block will be proposed, preconfs offered by MEV-Boost builders are conditional on preconfing builders winning the MEV-Boost auction~\cite{W:PreconfirmationFairExchange,W:LeaderlessandLeader-BasedPreconfirmations}. That being said,
        at the time of writing, three builders typically win more than 90\% MEV-Boost auctions \cite{MEV.pics}. Therefore, acquiring preconfs from these builders may be a sufficient commitment for some users and applications. 
    
\section{The Preconfirmation Pipeline}
\label{sec:pipeline}

\begin{figure*}[htbp]
        \centering
        \includegraphics[width=0.9\textwidth]{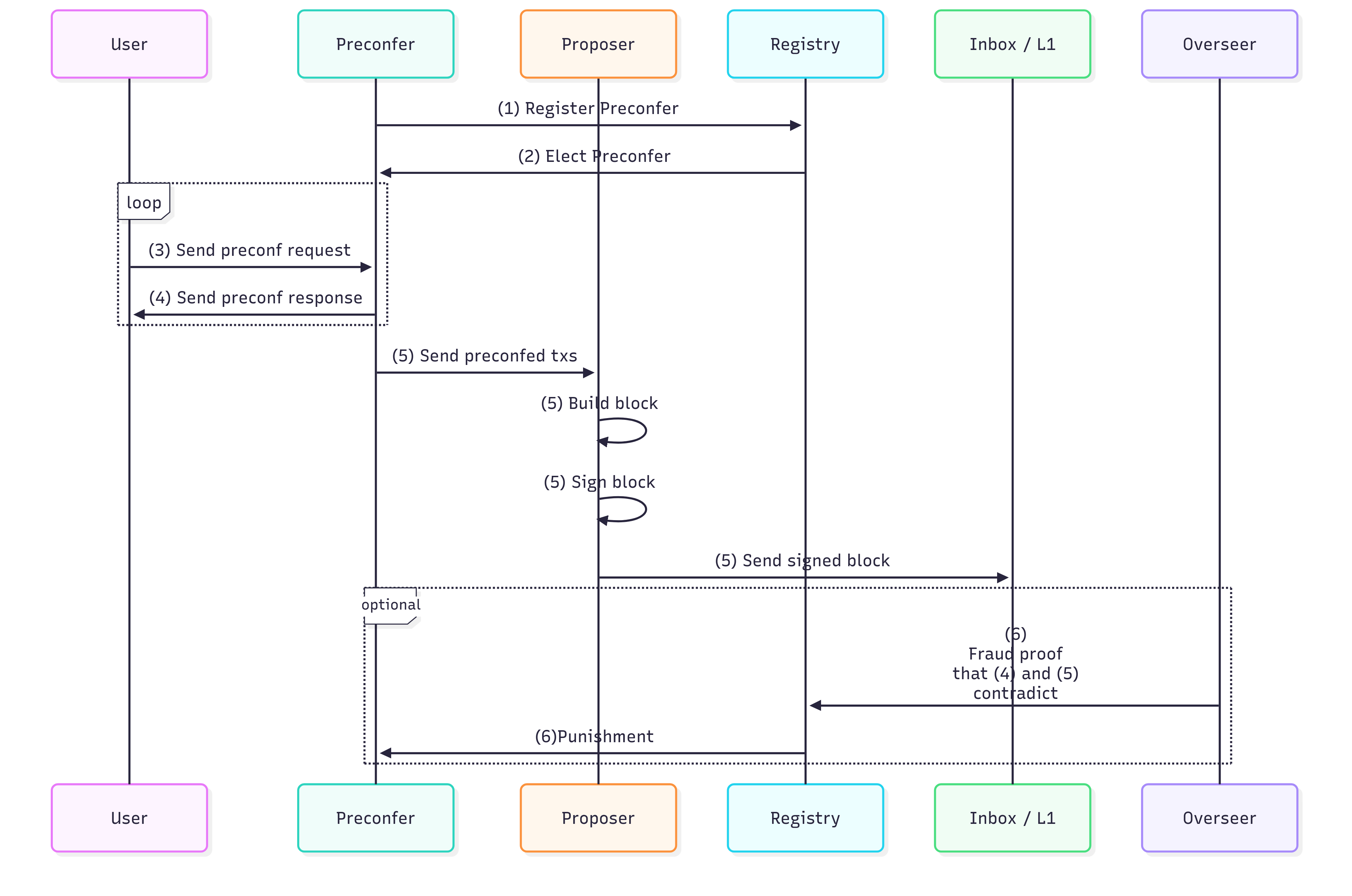}
        \caption{Preconf Protocol Flow.}
        \label{preconf_protocol_flow}
    \end{figure*}

This section analyzes the general flow of a preconf protocol and the roles of its key participants. The flow begins when a candidate expresses interest in joining the protocol as a preconfer. Upon meeting the protocol’s registration requirements, the candidate becomes eligible for election as a preconfer. The protocol then enters a repeated interaction phase involving users who submit preconf requests and preconfers who respond to them with preconf responses\footnote{In some preconf protocols, this phase may involve a trusted party, known as overseer, who signs the preconf responses along with the preconfer and penalizes them in real-time if the preconfer deviates from the protocol's instructions. This is a real-time punishment that is thoroughly discussed in \hyperref[punishments]{Section \ref{punishments}}.}. The final phase of preconf protocols focuses on the fulfillment of preconf promises and the on-chain publication (inclusion in the blockchain ledger) of any preconfed transactions. An optional addition to the flow concerns the punishment of preconfers who behave maliciously. \hyperref[preconf_protocol_flow]{Fig.~\ref{preconf_protocol_flow}} illustrates a general overview of the preconf protocol flow.

The remainder of this section goes through this preconf protocol flow, which we decompose into the six steps of preconfing.

\subsection{Step 1: Preconfer Registration} 
\label{step1:preconfer_registration}
    The first step in the preconf protocol flow is preconfer registration (see \hyperref[preconf_protocol_flow]{Fig.~\ref{preconf_protocol_flow}}). Preconfer registration involves candidates signaling that they want to become a preconfer. For this signaling to result in a successful registration, the candidate must meet all applicable eligibility criteria.
    This registration is typically managed by a smart contract known as the preconf registry, for which we give details in \hyperref[preconfer_registry]{Section~\ref{preconfer_registry}}.
    Each preconf registry has the freedom to set its own criteria for its preconfers. Some example criteria include:
    \begin{itemize}
        \item Proof that a registrant is a proposer. Generally, preconfers must have the ability to perform inclusion/execution of a preconfed transaction. Some preconf protocols may insist that preconfers be proposers of the underlying blockchain protocol. Specifically, registrants may need to prove:
        \begin{itemize}
            \item Membership of the L1 proposer set for L1 and based L2 preconfs.
            \item Governance committee approval for non-based L2 preconfs.
        \end{itemize}
        \item Deposit of some minimum collateral amount that may be subject to slashing~\cite{W:CrediblyNeutralPreconfirmationCollateral:ThePreconfirmationRegistry, W:PreconfirmationRegistry}.
        \item That a registrant is not blacklisted see \cite{W:PreconfirmationFairExchange}.     
    \end{itemize}

        \subsubsection{The Preconfer Registry} \label{preconfer_registry}
        In this section, we present what the smart contract preconf registry is usually responsible for, and we elaborate on the minimum collateral an entity may need to deposit to be registered as a preconfer.
        In \hyperref[sec:background]{Section \ref{sec:background}} we mentioned that every validator needs to deposit collateral to participate in block proposal and attestation in Ethereum. Here, we discuss what slashable collateral is in the context of preconf protocols. 
        \begin{definition}
            In the context of preconfs and a specific preconfer, \textbf{slashable collateral} is an amount of tokens that can be slashed in the case of that preconfer's misbehavior. 
        \end{definition}
        With the definition of slashable collateral in the context of preconf protocols established, we can now present the definition of the preconf registry.
        \begin{definition}
        \label{def: preconf_registry}
        A \textbf{preconfer registry} is a smart contract wherein entities can opt in to preconfing, registering themselves as preconfers. This smart contract can: (i) verify whether a candidate meets the applicable eligibility criteria; (ii) hold collateral and enforce monetary penalties on preconfers deemed to have acted maliciously in accordance with contract rules; and (iii) determine the preconfer schedule.
        \end{definition}

        Preconfer registries \cite{W:CrediblyNeutralPreconfirmationCollateral:ThePreconfirmationRegistry,W:UniversalRegistryContract} can play an important role in their respective blockchains. For L2s, preconfer registries can be tightly coupled with the proposer election mechanism, including having full control over proposer election (for more details, see \hyperref[preconf_election]{Section \ref{preconf_election}}).
 
        Preconfer registries where proposers can opt in to preconfing are expected to require preconfers to deposit some slashable collateral. In such protocols, the preconfer registry would be entrusted to manage the collateral on behalf of preconfers. This includes slashing of the collateral when deemed necessary by any registry-specified rules, known as slashing conditions\cite{W:Documentation-RegisteringasaProvider,W:PreconfirmationRegistry}. 
        These slashing conditions should in theory reflect the behaviours that are desired/acceptable from a preconfer. Slashing of collateral, and alternative punishment methods are thoroughly discussed in \hyperref[preconfer_punishment]{Section~\ref{preconfer_punishment}}. Note that in the case of conditional preconfs, the preconfer should only be eligible for penalization if the preconf condition is satisfied but the promised predicate is not satisfied (see \cite{W:Documentation-Understandingmev-commit} for more details on conditional preconf slashing). 
         
        Note that any preconfer collateral directly locked up in a preconfer registry cannot be reused for other purposes. Therefore, protocols may want to rely on third-party restaking platforms for sourcing slashable collateral, such as the EigenLayer platform~\cite{W:RestakingOverview}. Restaking allows collateral that secures the underlying blockchain to be reused for multiple purposes, making it an attractive and capital-efficient way to source collateral. Of course, repurposing slashable collateral for multiple protocols introduces risks related to stake being slashed in multiple protocols at once. Most, if not all collateral-dependent protocols design slashing rules assuming all of the deposited collateral is available for slashing. As such, if collateral in one protocol can be reduced without slashing or withdrawing from that protocol, the implied security of the collateral is weakened. For more information on the trade-offs of re-staking collateral, see \cite{W:Restaking101:APrimeronEigenLayer}.
        
        \par 
        
        Lastly, in an attempt to make preconfing accessible to solo-stakers with limited resources, collateral delegation has been proposed \cite{W:CrediblyNeutralPreconfirmationCollateral:ThePreconfirmationRegistry}. Collateral delegation allows delegates to delegate stake to a delegatee who wishes to act as a preconfer but does not possess the required collateral. In such a setting, if a delegatee violates a slashing condition, it would be the delegates' stake that would be slashed. Sharing of preconfer rewards would be one reason for delegating collateral, although delegation comes with significant risk for delegates (see Section~\ref{sec:risk})).

        \paragraph{Universal Registry Contract}
        At the time of writing and to the best of our knowledge, all preconf protocols implement and deploy their own standalone registry contract. The Universal Registry Contract  (URC)~\cite{W:UniversalRegistryContract,W:GitHub-UniversalRegistryContract} aims to address the fragmentation risks that standalone preconfer registries inherit. The URC aims to be a generic open-source contract that removes the need to create and maintain individual preconfer registry contracts for each preconf protocol. The URC's promise is that individual protocols can customize their slashing and collateral requirements outside of the core URC. The URC aims to standardize:
        \begin{itemize}
            \item Preconfer registration and collateral posting.
            \item How slashing condition violations are reported and enforced. Specifically, the URC defines a common interface and process for enforcing slashing, without tying itself to any specific slashing logic.
        \end{itemize}

        With a single shared preconf contract like the URC, there is an easier path for preconf adoption for all stakeholders. One preconfer registry simplifies the tasks of assessing and comparing preconf protocols, slashing conditions, preconf guarantees and preconf risks (see \cite{W:UnifiedPreconfRegistry} for more details on the risks and benefits of the URC specifically). 

\subsection{Step 2: Preconfer Election} 
\label{preconf_election}

Preconfer election is a crucial part of most preconf protocols.
This section examines how preconfers are elected.
As already mentioned, in many L2s, the preconfer registry and proposer election are aware of each other. In non-based L2s \cite{Optimism,ZKsync,Arbitrum}, the two tasks are handled by the same smart contracts. In contrast, for the L1 and based L2s, the proposer election mechanism is typically not aware of the preconf registry, or even that preconfs are taking place. In all cases, the preconfer registry must be consulted to identify the schedule of preconfers, and for which slots preconfs can be provided.
The preconfer schedule depends on (i) the type of the preconf (inclusion or execution) and (ii) how blockchain proposers delegate block-building. 

\paragraph{How the type of preconf affects the preconfer schedule}

\begin{itemize}
    \item \textbf{Execution preconfs:} If the current proposer of a blockchain is not a preconfer, then execution preconfs cannot be credibly provided, even if future proposers in the proposer lookahead are preconfers. This is because the current proposer has control over the next blockchain ledger update. As such, future preconfers cannot know the blockchain ledger on which they will act. Note that this does not apply to conditional preconfs that can be provided, for example, by a builder in MEV-Boost \cite{MEV-Boost}, even if the builder cannot be certain they will win the block-building auction.
    \item \textbf{Inclusion preconfs:}
     Such preconfs can be issued by many, if not all preconfers. This is because the fulfillment of inclusion does not depend on the current state of the blockchain ledger.  
\end{itemize}

\paragraph{How block-building delegation affects schedule}
\begin{itemize}
    \item \textbf{L1 and based L2s proposer preconfs:}  Proposers are determined via the L1 proposer election mechanism. L1 preconfer registries -- while not responsible for electing L1 proposers -- can still determine the preconfer schedule, e.g., through EIP-7917~\cite{EIP7917}. The preconfer schedule can be determined by finding the intersection of the registered preconfer set and the proposer lookahead, and taking into account whether the preconf is of type inclusion or execution. For based L2s, if the lookahead yields no eligible registered preconfer, a fallback mechanism (e.g., drawing from a static whitelist or randomly sampling from the registry) can be used to prevent gaps in preconf coverage (see \cite{fallback}).
    \item \textbf{Non-based L2s proposer preconfs:} Within non-based L2s, there is a large proposer-preconfer election design space. The preconfer registry can either read from the outputs of a standalone proposer election mechanism as in L1 and based L2 preconfs, or be integrated into the proposer election mechanism. 
     \item \textbf{Block builder preconfs:} When block builders issue preconfs, the preconfer schedule cannot be predetermined but rather depends on the outcome of the block-building auction. In this case, all preconf-registered block builders can issue conditional preconfs, conditioned on the corresponding block builder winning the block builder auction. Which of the builders wins the auction and thus is responsible for fulfilling the preconfs is then determined after preconfing has taken place. Given the probabilistic nature of block-builder preconfs being fulfilled, it may make sense for users to source conditional preconfs from more than one registered block builder~\cite{W:LeaderlessandLeader-BasedPreconfirmations}.
\end{itemize}

\subsection{Step 3: Preconf Request} \label{preconf_request}
    As soon as a registered preconfer is elected, users who wish to receive preconfs can begin submitting preconf requests (see \hyperref[preconf_protocol_flow]{Fig.~\ref{preconf_protocol_flow}}). Depending on the preconf protocol, users can route preconf requests to the preconfer via P2P~\cite{W:Documentation-Understandingmev-commit} or through RPC endpoints~\cite{W:Taiyi-off-chaincomponents,W:Towardsanimplementationofbasedpreconfirmationsleveragingrestaking}. The request structure can vary from protocol to protocol. Preconf requests may include the following:

    \begin{itemize}
        \item \textbf{Transaction(s):} Requests are expected to include the transaction(s) for which the user wishes to receive a preconf\footnote{Other preconf protocols allow users to request intent fulfillment \cite{W:Intent-BasedArchitectureandTheirRisks} or ahead-of-time commitments to include a transaction where the executable transaction has not been specified yet~\cite{W:Proposer-CommitmentInfrastructureinEthereum,W:OpportunitiesandConsiderationsofEthereumsBlockspaceFuture, W:BlockspaceFutures}. These latter preconfs are sometimes referred to as block-space commitments.}. 
        \item \textbf{Preconf tip:} The preconf tip is the compensation to be paid to the preconfer for the preconf. In some preconf protocols, the tip is atomically bundled with the transaction that requests a preconf. This means that the preconfer cannot redeem the tip without including the transaction in the blockchain ledger (for more details, see \cite{W:TheDerivationPipeline,W:AnalyzingBFTProposer-PromisedPreconfirmations}). An idealized tipping mechanism protocol should ensure that a preconfer is only rewarded with a preconf tip if both the preconf is fulfilled, and that preconfer is responsible for fulfilling the preconf. 
        
        Regardless of how the tipping mechanism is implemented, the preconf tip stands a key incentive for preconfers to provide preconfs. As such, pricing models for preconf tips are crucial. Preconf tip pricing is handled in detail in \hyperref[sec:price]{Section~\ref{sec:price}}.

        \item \textbf{Preconf type:} 
        Provided that the protocol and/or the preconfer support both inclusion and execution preconfs, users should specify which type of preconf they want.
    \end{itemize}

    Requests may also contain more nuanced parameters:
    \begin{itemize}
        \item \textbf{Deadline:} A deadline parameter sets the latest possible slot/time after which a preconf request should be considered void, and its transaction(s) no longer includable by the preconfer~\cite{W:PreconfirmationsforVanillaBasedRollups}.
        
        \item \textbf{Tip decay/escalator mechanism:} 
        The preconf tip may need to change during the lifetime of a preconf request to reflect the preconf requester's urgency for the preconf to be fulfilled. A tip decay or escalator mechanism can achieve this. Tip decay mechanisms can incentivize individual preconfers to respond in a timely fashion to maximise the tip revenue they receive\cite{W:Documentation-BidDecayMechanism}. Tip escalators have also been discussed for regular transactions in \cite{W:OrderflowauctionsandcentralisationII:orderflowauctions}. Tip escalators depend on multiple entities competing to fulfill a transaction or preconf first to collect the fee/tip. This makes such a tip escalator more appropriate for inclusion and conditional execution preconfs, where multiple preconfers may have the ability to fulfill a preconf.   
        
        \item \textbf{Preconf penalty:} Instead of relying on a predetermined penalty system, it is possible that users define penalties for preconfer faults related to a specific request through a request parameter~\cite{W:User-DefinedPenalties:EnsuringHonestPreconfBehavior, W:Documentation-BidStructure}. The idea is to allow users and preconfers to mutually agree on a level of cryptoeconomic security where a "one-size-fits-all" approach does not suffice.

        \item \textbf{Latest preconfed state:} 
        Protocols can allow users to specify a commitment to a transaction sequence that includes: (i) the transactions present in the blockchain ledger at the beginning of the current blockchain slot, and (ii) any transactions that have been preconfed. This gives users control over the transaction sequence their request acts on.  ~\cite{W:AnalyzingBFTProposer-PromisedPreconfirmations}

        \item \textbf{Privacy preferences:} 
        Privacy-preserving techniques may be applied to preconf requests to hide key request data from preconfers before they issue the preconf. 
        Techniques include public-key encryption, or trusted intermediaries as they used in MEV-Boost today~\cite{W:BasedPreconfirmationswithMulti-roundMEV-Boost, W:AnalyzingBFTProposer-PromisedPreconfirmations, W:Documentation-BidStructure, W:Towardsanimplementationofbasedpreconfirmationsleveragingrestaking}.
         
    \end{itemize}

\subsection{Step 4: Preconf Response} \label{preconf_response}
      
    After observing a preconf request and verifying its validity, the preconfer must then choose whether to provide a preconf or not. This section describes the process of responding to preconf requests. When a preconfer responds to a preconf request, the digital artifact sent to the user is the preconf. Generally, preconfers should be incentivized by the protocol to provide timely preconfs to users in order to improve user experience. To incentivize timely preconfs, a suitable tipping mechanism is required (as discussed in \hyperref[preconf_request]{Section \ref{preconf_request}} and \hyperref[sec:price]{Section \ref{sec:price}}), along with a protocol that ensures the preconfer receives the tip only if the preconf is provided no later than the agreed-upon time (as discussed in \hyperref[fair_exchange_problem]{Section  \ref{fair_exchange_problem}}).

    The response to a preconf request stands as a signed commitment to include or execute the preconfed transaction. In many preconf protocols, preconfs also serve as evidence for punishing dishonest preconfers who fail to fulfill stated commitments (e.g., in URC described in \hyperref[preconfer_registry]{Section \ref{preconfer_registry}}).
    Preconfs must be signed by the preconfer. Additionally, preconfs may include (but are not restricted to):
    \begin{itemize}
        \item \textbf{Unique identifier to a preconf request:}
        A preconf must contain a unique mapping to the request it is directed at. A hash of a signed request is such a mapping~\cite{W:Documentation-Commitments}.
        
        \item \textbf{Block number containing the requested transaction or validity period:} Some preconf protocols may require the preconfer to disclose the number of the block that will contain the requested transaction, or commit to the period during which the preconf will be fulfilled~\cite{W:Towardsanimplementationofbasedpreconfirmationsleveragingrestaking}.

        \item \textbf{Latest preconf state:}
        The preconfer may be required to provide a commitment to the  the latest preconfed state (explained in \hyperref[preconf_request]{Section \ref{preconf_request}}) at the instant when the preconf was provided. 
        
        \item \textbf{Commitment to additional slashing conditions:} Certain preconf protocols may require a preconfer to commit to additional slashing conditions -- e.g., as specified by the associated preconf request.
        \item \textbf{Timestamp:} The time at which the preconf was issued.
        
    \end{itemize}

    Apart from the content of a preconf, the frequency of preconfs can also vary. Preconfs can be published in several ways:
    \begin{itemize}
        \item Immediately upon commitment to fulfill the preconf, known as streaming (for example, as discussed in \cite{W:ThePreconfirmationGatewayUnlockingPreconfirmations:FromUsertoPreconfer}). Requiring the preconfer to publish a preconf before being allowed to continue preconfing could act as a disincentive to accept tips without providing timely preconfs. 
        
        \item At regular time intervals.
        \item In batches -- e.g., after the preconfer has committed to preconfing ten transactions, the preconfer publishes a preconf batch including these ten transactions.
        \item Upon request, granting access to previously issued preconfs.
    \end{itemize}

    The frequency with which preconfs are expected, enforced, and actually provided has implications for all entities in a preconf protocol. For example, streaming preconfs is resource intensive for a preconfer, both from an infrastructure and a pricing perspective (see \hyperref[sec:price]{Section \ref{sec:price}} for more details on how preconf frequency affects preconf pricing and preconfer revenue). On the other hand, streaming preconfs may offer certain users an improved user experience compared to batch-preconfing (see \cite{W:ThePreconfirmationGatewayUnlockingPreconfirmations:FromUsertoPreconfer}). In turn, this improved user experience could translate into increased preconf tips for the preconfer.

\subsubsection{Fair-Exchange of Preconf Requests and Responses} \label{fair_exchange_problem}
    Equally important to the core concepts of preconf requests and responses is the fair exchange of these requests and responses. 
    \begin{definition}
        
    The \textbf{fair exchange problem} states that during a mutual exchange of items between two entities, it is vital to guarantee that both entities or neither entity receive the expected item. No entity should have the ability to gain an advantage by misbehaving or quitting the transaction prematurely~\cite{P:Fairexchangewithasemi-trustedthirdparty,T:Fairnessinelectroniccommerce}.
    \end{definition}

    In the context of preconfs, users should have some expectation of how and when preconfs are received after submitting a request. This expectation depends on many factors, including but not limited to the type of preconf requested, the tip being paid, the cost to provide and fulfill the preconf, the relative demand for transactions/preconfs at the time the request was sent, and the value that a preconfer can extract from the preconf beyond the tip. Crucially, a preconf user's expectation has a time component built in -- users typically want preconfs sooner rather than later. This makes the fair exchange of preconfs more complex than eventual fair exchange of request and response. In \cite{W:PreconfirmationFairExchange}, the concept of timely fair exchange is introduced to explain this user preference in the context of preconf protocols. Previous work on fair exchange proved that it is impossible to enforce fair exchange without a trusted third party~\cite{P:OntheImpossibilityofFairExchangewithoutaTrustedThirdParty}. As such, preconf protocols need to introduce some form of trusted third party in order to provide meaningful guarantees of timely fair exchange. Note that a smart contract can serve as a trusted party -- assuming that the L1 satisfies certain security guarantees and the contract is correctly implemented -- but only in cases where there exists programmable proof of the preconfer’s misbehavior that the contract can independently verify. For example, the smart contract must have access to the proposer lookahead mechanism to confirm whether a specific entity signing a preconf was indeed elected as a preconfer for a given slot. 
    
    \begin{definition}
    \label{def:overseer}
    In the context of preconfs, an \textbf{overseer} is an entity or set of entities that is trusted to observe the actions of preconfers, and provide signals to the protocol in the case of a preconfer's deviation from protocol rules or expected actions. These signals can include proofs of deviation, although some deviations may not be provable, depending instead on overseer trust. Overseer signals may be communicated on-chain through protocol smart contracts, or off-chain through P2P communication layers.
    \end{definition}

    Several overseer election mechanisms have been discussed. The primary distinction of these protocols is based on whether the overseer can signal preconfer misbehavior in-protocol or not. In-protocol signaling allows for explicit preconfer punishment, something which is discussed in \hyperref[preconfer_punishment]{Section~\ref{preconfer_punishment}}.
    Such overseers can be elected through governance and/or be required to hold some large amount of a blockchain's native token for which the preconfs are being provided. Both of these mechanisms align overseers with the success and reliability of the preconf protocol. 
    
    For overseers who are expected to signal preconfer deviation out-of-protocol, several solutions have been discussed to date. In \cite{W:PreconfirmationFairExchange}, users are collectively identified as a suitable overseer in certain preconf protocols. Users can use the threat of withholding orderflow from a misbehaving preconfer to incentivize preconfers to follow protocol rules. The effectiveness of orderflow withholding as a threat depends on the expected profit of a preconfer from preconfs. This in turn is influenced by the frequency of being elected as a proposer/preconfer and the availability of block-space (see \cite{W:Future-ProofingPreconfirmations}).
    
    Wallet providers operating as out-of-protocol overseers on behalf of users is also discussed in \cite{W:PreconfirmationFairExchange}. As wallet providers are more sophisticated than a typical blockchain user, wallet providers may be better suited to identify preconfer deviations.
    In a similar vein, \cite{W:ThePreconfirmationGatewayUnlockingPreconfirmations:FromUsertoPreconfer} presents the concept of gateways (\hyperref[def:preconfirmation]{see Definition \ref{def:preconfirmation}}) and relayers (\hyperref[sec:L1_pipeline]{see Section \ref{sec:L1_pipeline}}) acting as overseers on behalf of users by monitoring preconfer behavior and withholding requests or tips. It is important to note that overseers and preconfers should be distinct entities in each preconf protocol. For example, a relayer can assume the roles of a preconfing gateway and an overseer, but not in the same preconf protocol.  

\subsection{Step 5: Fulfillment }\label{preconf_delivery}

It is a preconfer's ability to fulfill a preconf that provides a preconf protocol with utility. This section focuses on the different mechanisms that preconfers can use to fulfill preconfs. These fulfillment mechanisms depend on many factors, including but not limited to:
\begin{itemize}
    \item What layer of blockchain the preconfs apply to.
    \item How much of a block is being preconfed, and how any remaining block-space is filled when only a percentage of a block is being used for preconfs.
    \item Whether the preconfer is a proposer or gateway.
\end{itemize} 
These tradeoffs are examined in the following Fulfillment Tradeoffs subsections.

\subsubsection{Fulfillment Tradeoffs: Based vs Non-Based}

For L1 and based L2 preconfs, the L1 proposer controls proposal of L1 blocks, and so can always propose an L1 block containing any and all preconfs. For non-based L2s with a single proposer, fulfillment is also controlled by the proposer themselves. 

In the case of non-based L2s with mutliple proposers, the rules of the underlying L2's blockchain ledger dictate a preconfer's ability to fulfill preconfs. In some non-based L2s, it may be sufficient to just propagate the block to the P2P layer for consensus. However, some non-based L2s require that an L2 proposer's blocks must appear in the L1 ledger by a certain proposal deadline to be considered valid. In this case, an L2 preconfer can only fulfill their preconfs if an L1 proposer includes the corresponding L2 blocks on L1 by the proposal deadline, which is not guaranteed. 

\subsubsection{Fulfillment Tradeoffs: Full Block Preconfs vs Partial Block Preconfs}

In the case where a full block is being used for preconfs, a preconfer can build the block themselves and ensure fulfillment of all preconfs. In the case where only a partial block is used for preconfs, and the final block is built at proposal time, potentially including regular transactions, there must be a mechanism in place to ensure the final block fulfills all preconfs. A simple algorithm for achieving this is including all preconfed transactions first in a block, followed by the remainder of the block. However, when any of these preconfed transactions are inclusion preconfs, this simple algorithm is likely sub-optimal for a proposer with regard to capturing block-building revenue. As inclusion preconfs can be placed anywhere in a block, a proposer can strategically place these preconfs to maximize their MEV revenue. 

That being said, this problem of maximizing the value of blocks given some constraints has parallels to the ordered knapsack problem ~\cite{P:Partiallyorderedknapsackandapplicationstoscheculing}, which requires significant sophistication to solve for the knapsack sizes and number of items that modern blockchains provide. This problem has been referred to as the Online Block Packing problem \cite{OnlineBlockPacking}. To allow proposers to outsource this complex optimization problem while still receiving revenue from the solution, the Constraints API~\cite{W:ConstraintsAPISpecification} has been proposed as a tool to allow proposers to offer preconfs (or apply arbitrary constraints on a block), and then outsource the building of a block satisfying all preconf predicates to a set of builders, analogously to how proposers outsource block-building through MEV-Boost \cite{MEV-Boost}. 

\subsubsection{Fulfillment Tradeoffs: Proposer vs Gateway}
\label{delivery_tradeoffs}

Proposers are well placed to provide preconfs, since they can include them directly in their own blocks—unlike gateways.  For gateway preconfs, the proposer–gateway relationship is key: if the gateway trusts the proposer, straightforward message-passing algorithm  suffices; otherwise, cryptoeconomic guarantees are needed to ensure all gateway-issued preconfs are included, especially if the gateway is liable to be punished for failures. Such a guarantee could include shared punishment when preconfs are not fulfilled, or the existence of a mutually trusted overseer to arbitrate faults \cite{W:PreconfirmationFairExchange,W:FaultAttribution}. 
Depending on whether or not the gateway is also tasked with sourcing the block for the proposer will dictate when or if the preconfs must be sent back to the proposer. In the case where the gateway also sources the proposer's block, for example, if the gateway also runs Constraints API~\cite{W:ConstraintsAPISpecification} or Commit-Boost \cite{CommitBoostRepo}, the proposer can simply sign the block header and have the gateway propagate the final block to the P2P network, as in MEV-Boost.
\subsubsection{Designing Robust Fulfillment Mechanisms}\label{robust_delivery}

    In any preconf protocol, it is possible that a preconfer fails to fulfill their preconfs. Although this non-fulfillment may be malicious, non-fulfillment of preconfs can also be unintentional. Some thought has been given to protecting against unintentional failures to fulfill preconfs. The authors of \cite{W:AvoidingAccidentalLivenessFaultsforBasedPreconfs} suggest that preconfers can collaborate with subsequent preconfers through revenue sharing to preconf the same transaction sequences in the event that the original preconfer fails to fulfill the preconfs themselves. This approach is termed \textbf{chaining} preconfs. Although the exact incentives and guarantees of chaining are unexplored, chaining provides a protocol design direction to protect preconfers and their users from the risks of failing to fulfill preconfs.

\subsection{Step 6: Preconfer Faults \& Enforcement Mechanisms} \label{preconfer_punishment}

    Preconfer faults are part-and-parcel of preconfing, standing in the way of the ideal notion of perfectly-reliable preconfs. Preconf enforcement mechanisms are an important tool in disincentivizing faults and ensuring that preconfers either fulfill preconfs according to any preconf restrictions, or are punished. 
    
    \subsubsection{Preconfer Faults}\label{preconfer_faults_and_punishing_conditions}
\begin{figure}[htbp]
        \centering
        \includegraphics[width=0.9\textwidth]{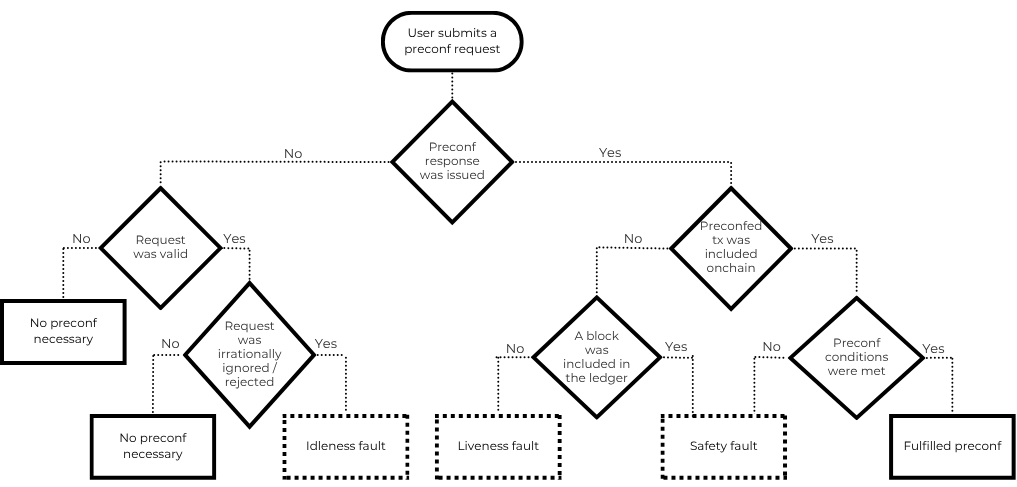}
        \caption{Preconfer Fault Decision Tree}
        \label{preconfer fault tree}
    \end{figure}

    There are various faults that a malicious or negligent preconfer can commit that jeopardize the smooth operation of a preconf protocol. \hyperref[preconfer fault tree]{Fig.~\ref{preconfer fault tree}} presents a decision tree that categorizes preconfer faults. These preconfer faults are described in the remainder of this subsection. 

        \begin{definition}
            A \textbf{safety fault} occurs when the proposer publishes a block which violates a preconf predicate to which the preconfer committed.
        \end{definition}
        
    \par
    Note that a safety fault will never be committed by an honest preconfer who is also a proposer. In such cases, the punishment is expected to be severe~\cite{W:Basedpreconfirmations}.
    
    Unlike safety faults, preconf liveness faults do no clearly apply to all blockchains. We define liveness faults for L1s and based L2s in-line with the seminal work on based preconfirmations \cite{W:Basedpreconfirmations}.
    
        \begin{definition}
            \label{def:liveness_fault}
            A \textbf{liveness fault for L1 and based L2 preconfs} occurs when an L1 proposer does not publish an L1 block for a particular slot, meaning all preconfed transactions for that slot are not included in the respective blockchain ledger.
        \end{definition}

         If we attempt to apply a similar definition to non-based L2s, preconf liveness faults are in many cases equivalent to a liveness failure of the underlying L2 -- which is conceptually different to the well-understood liveness fault of L1s and based L2s. As such, we do not consider liveness faults with respect to non-based L2s. 
         
         Unlike safety faults, liveness faults can be accidental and non-malicious, for example, due to a power outage or internet downtime. Therefore, preconf protocols may prefer a lighter penalty for liveness faults, as proposed in~\cite{W:Basedpreconfirmations}, or to introduce fulfillment mechanisms that mitigate the risk of preconfers being slashed for accidental violations, as discussed in \hyperref[robust_delivery]{Section~\ref{robust_delivery}}.     

     Although idleness faults are not well-documented, it is natural for some preconf protocols to require and enforce that preconfers issue preconfs under certain conditions, e.g., preconfing all valid requests, or preconfing all requests paying tips greater than a certain amount.
        \begin{definition}
            An \textbf{idleness fault} occurs when a preconfer receives valid preconf requests but neglects their duty and fails to provide preconf services.
        \end{definition}

        As illustrated in \hyperref[preconf_response]{Section~\ref{preconf_response}} and \hyperref[sec:price]{Section~\ref{sec:price}}, there are various factors that can influence a preconfer's decision to provide a preconf, such as the preconf tip and the incompatibility of conflicting transactions. As idleness faults depend on requests being delivered to the preconfer but not being responded to, this introduces a subjective observability requirement that could potentially be handled by a trusted party such as an overseer as presented in \hyperref[fair_exchange_problem]{Section \ref{fair_exchange_problem}}, in \hyperref[def:overseer]{Definition \ref{def:overseer}}.
        

\subsubsection{Enforcement Mechanisms and Proofs of Malicious Behaviour}
\label{sec:enforcement}

    Preconf protocols rely on some form of economic incentive to encourage honest preconfer behaviors. These incentives are provided by mechanisms that either explicitly or implicitly punish malicious preconfer behavior. 
    \begin{definition}
    \label{def:enforcement}
        An \textbf{enforcement mechanism} 
        triggers punishment of a preconfer when the preconfer satisfies one or more punishing conditions. 
    \end{definition}

    There are many possible enforcement mechanism implementations. Key differences between enforcement mechanisms include whether the overseer responsible for observing punishing conditions is permissionless or permissioned, the type of proof required to trigger preconfer punishment, and the level of punishment~\cite{W:PreconfirmationFairExchange}.

    In a permissioned overseer setting, observation and triggering of punishments are driven by the permissioned overseer. Generally, the proofs required to initiate punishment for permissioned overseers are less rigorous.  Punishing conditions do not necessarily need to be proven to be satisfied, potentially only requiring an overseer signature. When proofs are not required to trigger punishments, this raises concerns about overseer abuse~\cite{W:PreconfirmationFairExchange}.

    In a permissionless setting, anyone can act as an overseer to initiate punishment via an enforcement mechanism. Contrary to the permissioned setting, permissionless overseers must be required to submit conclusive and indisputable evidence to trigger explicit preconfer punishments via the enforcement mechanism. Such evidence may include a signed preconf request, response pair disagreeing with the blockchain ledger. When indisputable evidence is not available, some form of arbitration between preconfer and overseer(s) must take place. As triggers are disputable in this case, this exposes permissionless overseer enforcement mechanisms to incorrect or unsatisfactory triggering and punishment~\cite{W:GitHub-UniversalRegistryContract,W:GitHub-ExampleSlasherImplementations,W:PreconfirmationFairExchange}. 
    
    Another important aspect related to the indisputable evidence required to trigger a preconf punishment is computational complexity. The authors in \cite{W:ATaxonomyofPreconfirmationGuaranteesandTheirSlashingConditionsinRollups} discuss how this complexity depends on the type of preconf -- inclusion versus execution. For execution preconfs, it is sufficient to prove that, at the specific position where the preconfed transaction is expected to appear, a different transaction exists. In contrast, for inclusion preconfs, the overseer must prove that a preconfed transaction is absent from every position in all blocks proposed after when the preconf was given until any imposed deadline for fulfillment. This deadline can be protocol-imposed, or imposed by the request (see Section \ref{preconf_request}).

    \subsubsection{Punishments} \label{punishments}

\begin{table}[h]
    \renewcommand{\arraystretch}{1.5} 
    \centering
    \begin{tabularx}{\textwidth}{| l | X | X |}
        \hline
        & \textbf{Direct} & \textbf{Indirect} \\
        \hline
        \textbf{Real-time} & Overseer interrupts preconfs or proposals & Downgrade reputation to affect orderflow \\
        \hline
        \multirow{3}{*}{\textbf{Ex-post}} 
        & Slashing & Downgrade reputation to affect orderflow \\
        \cline{2-3}
        & Temporary black-listing & Forfeit revenue while blacklisted \\
        \cline{2-3} 
        & Temporary stake freezing & \\
        \hline
    \end{tabularx}
    \caption{Preconfer Punishment Mechanisms}
    \label{tab:punishments}
\end{table}

    As mentioned in the previous section, preconfer punishments can be divided into \textbf{direct} and \textbf{indirect} punishments, depending on how punishments are applied to a preconfer's net worth. Direct preconfer punishments involve preconfer collateral slashing, while indirect preconfer punishments target revenue from preconfs~\cite{W:PreconfirmationFairExchange}. For example, if a preconfer is punished through blacklisting~\cite{W:PreconfirmationFairExchange} for a specific period, the preconfer forfeits all revenue from preconfs during that time.
    
    Preconfer punishments can also be divided into \textbf{real-time} and \textbf{ex-post} punishments, depending on when punishments are applied \cite{W:PreconfirmationFairExchange}. In real-time punishment, preconfer's actions are monitored by an overseer and any misbehaviour is immediately detected preventing the preconfer from making any profit. For instance, in real-time punishments, the preconf responses may also need to be signed by the overseer. This means that the issuance process can be disrupted not only when the preconfer is idle, but also if the overseer fails to perform their role. This creates a dependency on overseer liveness for the preconf protocol to function.
    
    Ex-post punishments penalize dishonest preconfers retroactively. This removes the preconfer as a liveness dependency for the preconf protocol. However, ex-post punishments typically require preconfers to lock up collateral that can be slashed. If any of the slashing conditions are violated, the preconfer loses a predetermined amount of their collateral. Moreover, a preconfer can be temporarily blacklisted, i.e., be deprived of their right to provide preconfs or have their stake temporarily frozen. In addition to these explicit punishments, enforcement mechanisms can also limit the preconf orderflow of dishonest preconfers or downgrade their reputation to affect future orderflow~\cite{W:PreconfirmationFairExchange,W:User-DefinedPenalties:EnsuringHonestPreconfBehavior}, implicitly reducing preconf revenue.

\section{Preconf Economics}
\label{preconf_economics}

This section focuses on preconf economics, with particular focus on how the revenue of proposers is expected to change given preconfs, how preconfs are priced, and the general economic viability of preconfs. Although the enforcement mechanisms described in \hyperref[punishments]{Section \ref{punishments}} fall under the heading of preconf economics, we omit any additional discussion on punishments in this section. 

By construction, preconfs require proposers to commit to block-space earlier than block proposal time. However, without enforcement of preconfs or explicit preconf rewards, preconfers have an incentive to withhold preconf responses as long as possible. All else equal, delaying the issuance of preconfs allows proposers to build blocks using up-to-date information, with the number of transactions in the mempool and number of possible transaction orderings increasing with time. These are all factors that contribute to a proposer's ability to extract MEV. 

In contrast, user demand for preconfs translates to users gaining utility by receiving a confirmation of transaction execution or inclusion before a proposer proposes a block, as already mentioned in \hyperref[fair_exchange_problem]{Section \ref{fair_exchange_problem}}.

To address the tension of users wanting preconfs, and proposers wanting to wait as long as possible before a response is provided, proposer incentives to provide preconfs are needed. We have already seen in \hyperref[preconfer_punishment]{Section~\ref{preconfer_punishment}} that the preconfer role can be subjected to punishment conditions to incentivize certain behaviours, including timely responses to requests. Just as punishments can be applied for slow responses, incentives in the form of preconf tips can be provided for timely responses. 

The remainder of this section considers how preconfs affect proposer revenues, and how preconf tips play an important role in the adoption of preconfs. We conclude the section by describing several preconf tip pricing models that have been discussed in the preconf literature. Equipped with well-founded preconf tip pricing models, users, proposers, and preconfers alike can use or adapt these models as required, and make informed decisions about preconf tip pricing.

\subsection{Impact of Preconfs on Proposer Revenue}
\label{preconfs_impact_proposer_revenue}
    
     Several works have attempted to model the effect of preconfs on expected proposer revenue \cite{W:PreconfirmationsundertheNOlens,W:EstimatingtheRevenuefromIndependentSub-SlotAuctionPreconfirmations,W:AnalysingExpectedProposerRevenuefromPreconfirmations,W:APricingModelforInclusionPreconfirmations, W:MeasuringValidatorEconomics}. Generally, preconfs increase demand for block-space by offering users an additional mechanism to express that demand. Unfortunately for proposers, block-space demand does not directly translate to increased proposer rewards. This is because many blockchains, including Ethereum and its L2s, deterministically increase transaction base fees (fees that go to the blockchain protocol and not to the proposer)~\cite{EIP1559_Economic_Analysis} when demand is high. Therefore, sustained demand resulting in increased base fees reduces the relative priority fee/tip that a proposer can capture. Demand aside, there are two key takeaways from existing literature on proposer revenue from preconfs: 
        
     \begin{enumerate}
         \item Inclusion preconfs are a relatively straightforward way for proposers to increase their revenue from proposing without disrupting their expected revenue from existing block-building strategies, such as MEV-Boost, commit boost, or otherwise \cite{W:APricingModelforInclusionPreconfirmations,W:PricingTransactionsforPreconfirmation}.
         \item Execution preconfs require preconfer sophistication and preconf-specific tip revenue to make execution preconfs economically viable for proposers with the option not to preconf \cite{W:PreconfirmationsundertheNOlens,W:EstimatingtheRevenuefromIndependentSub-SlotAuctionPreconfirmations,W:AnalysingExpectedProposerRevenuefromPreconfirmations}. 
     \end{enumerate}

    In the following subsections, we provide a detailed breakdown of how these takeaways are established, first for inclusion preconfs and then for execution preconfs. 
    
    \subsubsection{Inclusion Preconfs \& Proposer Revenue}
\label{inclusion_revenue}
    Inclusion preconfs provide proposers with a potential source of revenue through preconf tips without a clear need for increased proposer sophistication
    \cite{W:APricingModelforInclusionPreconfirmations, W:PricingTransactionsforPreconfirmation, W:MeasuringValidatorEconomics}. The study provided in~\cite{W:MeasuringValidatorEconomics}, although based on limited mainnet data from different preconf protocols, supports the thesis that preconfs add value to typical Ethereum blocks.

    In \cite{W:APricingModelforInclusionPreconfirmations, W:PricingTransactionsforPreconfirmation}, it is identified that proposers can use even basic preconf pricing models to offer preconfs and expect to increase their overall revenue. The key factors at play here are:
    \begin{enumerate}
        \item \textbf{Flexibility:} Inclusion preconfs give proposers the ability to choose where preconfed transactions are included in a block. Specifically, a transaction receiving a preconf can be placed anywhere in a block by a proposer.
        \item \textbf{Ease of Pricing:} The further down in the transaction list of a block a transaction 

        is positioned, the less value and volatility its expected fee (or preconf tip)
        has in dollar terms. These low price, low price-volatility parts of the blocks are where inclusion preconfs are expected to compete for inclusion in the blockchain ledger, given the flexibility that inclusion preconfs give to propsoer with respect to transaction positioning in a block. 

        This makes the pricing of inclusion preconfs straightforward for a proposer, as the expected error of pricing models is low. For example, a preconfer who knows that 50\% of transactions pay less than 1\$ per 100,000 gas used can happily sell 100,000 gas inclusion preconfs for 1\$ or more until the total gas consumed by inclusion preconfs is more than 50\% of the available block-space (gas represents a unit of computation, and each Ethereum block has a limit on the total amount of gas that can be consumed by its transactions). 
        \item \textbf{Compatibility with Block-Building Auctions:} Thanks to designs like Commit-Boost \cite{CommitBoostRepo} and the Constraints API \cite{W:ConstraintsAPISpecification}, any preconfs committed to by a preconfer can be communicated to and enforced upon builders. These implementations are practical implementations of earlier theoretical designs \cite{ResistanceisnotfutileCRinMEVBOOST}. Such designs allow proposers to benefit from MEV-Boost style competition (see PBS in \hyperref[sec:background]{Section \ref{sec:L1_pipeline}}) among builders on the remaining parts of the block, enhancing proposers' revenue. 
    \end{enumerate}
    Inclusion preconfs therefore stand as an easy-to-price revenue add-on for proposers. Of course, pricing sophistication is always possible. Proposers can likely increase expected revenue by using 
    more sophisticated pricing models compared to the conservative inclusion preconf pricing models presented here. However, expected or required sophistication is not without risks (see Section \ref{sec:risk}). 

    A caveat to using models that estimate preconf revenue by extrapolating regular transaction revenues, as is done in \cite{W:APricingModelforInclusionPreconfirmations} and \cite{W:PricingTransactionsforPreconfirmation}, is provided in~\cite{W:MeasuringValidatorEconomics}. The authors point out that one must be careful when looking at simple revenue averages since they can be heavily skewed by large outlier blocks with very high rewards. Although outliers are explicitly removed in both \cite{W:APricingModelforInclusionPreconfirmations} and \cite{W:PricingTransactionsforPreconfirmation}, this warning remains relevant to anyone looking to use such pricing models.

    \subsubsection{Execution Preconfs \& Proposer Revenue}\label{sec:execpreconfproposerrevenue}
    Unlike inclusion preconfs, execution preconfs must be executed in a specific way, typically at a specific position in a specific sequence of transactions. This restriction has implications on a proposer's expected revenue when offering execution preconfs. Although the proposer is acting as a preconfer, we use the term proposer in-keeping with the purpose of this section -- analysing the effect of preconfs on the proposer's revenue.  
    The models presented in \cite{W:EstimatingtheRevenuefromIndependentSub-SlotAuctionPreconfirmations}, \cite{ W:AnalysingExpectedProposerRevenuefromPreconfirmations}, and \cite{W:PreconfirmationsundertheNOlens} represent the main contributions in the space at time of writing. 
    
    The models in \cite{W:EstimatingtheRevenuefromIndependentSub-SlotAuctionPreconfirmations} and  \cite{W:AnalysingExpectedProposerRevenuefromPreconfirmations} are presented by the same set of authors. In these works, the authors claim that execution preconf protocols can be classified under two general classes of preconf protocols.  
    These classes are independent-sub-slot auctions (ISSAs)  \cite{W:EstimatingtheRevenuefromIndependentSub-SlotAuctionPreconfirmations} and dependent-sub-slot auctions (DSSAs) \cite{W:AnalysingExpectedProposerRevenuefromPreconfirmations}. In-keeping with the language of these works, we will refer to At a high-level, these 
    classes are described as follows:
    \begin{itemize}
        \item \textbf{Independent sub-slot auctions (ISSAs):} At its core, ISSAs propose proposers splitting slots into sub-slots and running independent MEV-Boost-style auctions to build sub-blocks at each sub-slot.  
        Each of the winning sub-blocks is signed by the proposer, and stand as the preconfs for that sub-slot. Given these preconfed sub-blocks, the final proposed block is the concatenation of each sub-block's transaction lists. The crucial property of ISSAs in contrast to DSSAs is the independence of the sub-slot auctions, where at each sub-slot, the proposer is trying to maximize the revenue from the current sub-slot auction. The ISSA framing is aligned with multi-round MEV-Boost~\cite{W:BasedPreconfirmationswithMulti-roundMEV-Boost}, where it is also suggested to run independent MEV-Boost-style auctions to build sub-blocks at each sub-slot.
        
        \item \textbf{Dependent sub-slot auctions (DSSAs):} As in ISSAs, DSSAs split a proposer's slot into sub-slots of arbitrary length, where sub-blocks are built at each sub-slot with the final proposed block being the concatenation of all sub-blocks. The distinction between DSSAs and ISSAs is that, at each sub-slot in DSSAs, the proposer chooses the sub-block which maximizes the expected revenue of the entire slot. This expected revenue includes the transaction fees from the block, preconf tips, and any MEV that can be captured, as well as the expected revenue of future sub-blocks.  
        
        Choosing sub-blocks that maximize a slot's expected revenue is an NP-Hard packing problem (think Knapsack Problem where the value of the contents and the size of the Knapsack are continuously changing) for proposers. As such, there is a high degree of proposer sophistication required to run a DSSA. 
    \end{itemize}

   Under this classification, an execution preconf protocol being an ISSA or DSSA has significant impact on expected proposer revenues. The main results on ISSAs and DSSAs are summarized as follows: 
    

    \begin{itemize}
        \item \textbf{ISSAs:} It is estimated in \cite{W:EstimatingtheRevenuefromIndependentSub-SlotAuctionPreconfirmations} that, \textit{ignoring any expected preconf tips}, preconfing using ISSAs with eight equal length sub-slots reduces a proposer's expected rewards for a slot by 50\% compared to a single end-of-slot MEV-Boost auction. At the limit, it is estimated that ISSAs with infinitesimally short sub-slots reduce a proposer's expected revenue by 74\% compared to a single end-of-slot MEV-Boost auction, \textit{ignoring any expected preconf tips}.

        \item \textbf{DSSAs:} In \cite{W:AnalysingExpectedProposerRevenuefromPreconfirmations}, it is proven that, \textit{ignoring any expected preconf tips}, a proposer's expected revenue through DSSAs is greater than or equal to the expected revenue from a single end-of-slot MEV-Boost auction. This result is derived by observing that a proposer's expected revenue from preconfing nothing at each sub-slot must be equal to that of running end-of-slot MEV-Boost. Therefore, any alternative maximization strategy chosen by 
        a proposer must have increased the expected proposer revenue. 
    \end{itemize}

    It is important to emphasize that these initial results exclude expected preconf tips from their analyses. Although the expected revenues from ISSAs appear pessimistic in comparison to DSSAs, it can be argued that the preconf tips in ISSAs are expected to be higher than the tips in DSSA\cite{W:AnalysingExpectedProposerRevenuefromPreconfirmations}. The reasoning here is that ISSAs enable and incentivize users to pay to update the entire blockchain state, including contentious state (see \hyperref[def:contentious state]{Definition \ref{def:contentious state}}) at each sub-slot. This is in comparison to DSSAs, where the proposer/preconfer on behalf of the proposer is expected to strategically delay updates to contentious state in order to maximize the full slot's revenue. The effect of proposers preferring to delay contentious state updates to capture more MEV has been formally proven to exist \cite{LVRwithFees}.

    In \cite{LVRwithFees}, it is proven that the value lost by liquidity providers in decentralized exchanges is strictly increasing in the time between state updates. As mentioned previously, ISSAs decrease the time between state updates compared to the normal block-building pipeline or DSSAs. In turn, liquidity providers under ISSAs would lose less money, meaning lower fee requirements and more liquidity for users. This creates a fly-wheel where users would be incentivized to trade more with decentralized exchanges, in turn paying more transaction fees and preconf tips.

    The simulation-based conclusions from \cite{W:PreconfirmationsundertheNOlens} are in line with the more formal results of \cite{W:AnalysingExpectedProposerRevenuefromPreconfirmations}. Both articles identify that proposers who consider the problem of building sub-blocks to maximize the expected revenue of the entire block stand to profit from preconfs. In \cite{W:PreconfirmationsundertheNOlens}, the authors argue that proposers  are likely to ignore preconfs on contentious state unless there are clear overpayments to act on this contentious state. This is because the sophisticated entities that normally pay most to act on contentious state pay more as slots progress, in line with the results from \cite{LVRwithFees}. It is again noted that despite not preconfing on contentious state, proposers can increase their expected revenue by preconfing on uncontentious state. The authors conclude by acknowledging the significant increase in sophistication required to preconf, whether through differentiating between contentious and uncontentious state, or determining over-payments on a per-transaction basis.  
    
    \subsection{Preconf Tip Pricing}
    \label{sec:price}
    
     Preconf tips are a critical component in any preconf protocol, in the same way that transaction fees are critical for blockchains. Preconf tips provide an incentive for preconfers to preconf a request, allowing users to express their preference for how quickly a request should be responded to (see \hyperref[preconf_request]{Section \ref{preconf_request}}). Therefore, the pricing of preconf tips to strike the balance between user and preconfer preferences to minimize costs yet maximize utility stands as an important area of research and development. Preconf pricing models vary in importance depending on whether or not the preconfer is trusted. 
     
     Preconfer trust can be as a result of an overseer enforcing preconfs (see \hyperref[sec:enforcement]{Section \ref{sec:enforcement}}) and punishing preconfers (see \hyperref[sec:enforcement]{Section \ref{punishments}}) for not preconfing fee-paying requests, or where the preconfers themselves are trusted, as in non-based single-proposer L2s \cite{Optimism,Arbitrum,ZKsync}. In such trusted preconfer protocols, preconfers are expected to preconf any request paying a tip above some protocol-specified threshold.
     This threshold is sometimes referred to as the L2 base fee. This L2 base fee is intended to cover an L2 transaction's costs, such as the costs of posting the transaction to the respective data-availability layer, or proving the transaction as part of a batch, where applicable. In systems where preconfs are expected to be given regardless of the preconf tip, preconf tip pricing is not meaningful.
     
     Preconf tip pricing becomes relevant for untrusted preconfer protocols, where the preconfer has freedom to respond positively or negatively to requests. In such settings, one must instead consider preconfers as trying to maximize the revenue of their entire preconf slot e.g., through MEV extraction. Maximizing revenue requires a preconfer to understand the value being forfeited when providing a preconf. Understanding this forfeited value, in theory, allows a preconfer to accept any preconf tip paying more than this amount. Although there is no established pricing method for preconfs, there are several resources that provide information on how preconfs can be priced~\cite{W:APricingModelforInclusionPreconfirmations,W:PricingTransactionsforPreconfirmation}. Pricing methodologies can be separated into those for inclusion preconfs~\cite{W:APricingModelforInclusionPreconfirmations,W:PricingTransactionsforPreconfirmation}, and those for execution preconfs, where no formal model for pricing has been established to the best our knowledge.
     
     \subsubsection{Inclusion Preconf Pricing}
     For inclusion preconf pricing, several models have been put forward~\cite{W:APricingModelforInclusionPreconfirmations,W:PricingTransactionsforPreconfirmation}. In \cite{W:APricingModelforInclusionPreconfirmations}, it is identified that block-space demand, measured in ETH per block-gas used, is highest for the first transactions in the transaction list of a block, and least valuable at the end of a block. Specifically, cumulative transaction fees in block-space follow a lognormal distribution. This aligns with the intuition that first access to contentious state has value for MEV extractors (see \hyperref[def:contentious state]{Definition \ref{def:contentious state}}).
     
     As already mentioned in \hyperref[inclusion_revenue]{Section \ref{inclusion_revenue}}, given that inclusion preconfs need only compete for the least valuable block-space, inclusion preconfs can be priced in an almost identical way to this least valuable block-space. Given the logarithmic nature of cumulative transaction fees versus block-space, \cite{W:APricingModelforInclusionPreconfirmations} demonstrates that basic price curves can be utilized by proposers to price inclusion preconfs for any amount of block-space. As an extension of this pricing curve,  \cite{W:APricingModelforInclusionPreconfirmations} identifies a pricing surface to incorporate both the amount of gas to be used by the preconf, and the amount of gas already consumed in the block. 
     
    The approach of \cite{W:PricingTransactionsforPreconfirmation} follows a similar methodology to \cite{W:APricingModelforInclusionPreconfirmations}, suggesting the use of historical transaction fees to price inclusion preconf tips. Rather than pricing inclusion preconfs using historical data for an entire block, \cite{W:PricingTransactionsforPreconfirmation} models inclusion preconf tips using transactions fees from the middle of blocks' transaction lists. This subset of transactions represents transactions paying some premium without competing for contentious state, which, as the authors argue, represents a similar class of transaction orderflow as inclusion preconfs. Both articles identify a need to increase preconf tips as more preconfs are provided and remaining block-space declines. 
    
    The models created in both \cite{W:APricingModelforInclusionPreconfirmations} and \cite{W:PricingTransactionsforPreconfirmation} leave room for refinement and improvement, albeit with the trade-off of increased preconfer sophistication requirements. Possible model adaptions can incorporate up-to-date estimates of block-building revenue or more granular per-unit block-space pricing models. 

    \subsubsection{Execution Preconf Pricing}

    To the best of our knowledge, there have been no meaningful pricing models proposed for execution preconfs. For a preconfer, pricing execution preconfs translates to continuously observing and analyzing incoming preconf requests throughout a slot. This is more computationally complex than building a single whole block at the end of a slot. Given the oligopolistic nature of MEV-Boost auctions \cite{MEVBoostShares} to build these whole blocks, the more complex task of execution preconfing and execution preconf pricing stands as a hard open problem for preconfs. 
  
    An indicator of this additional complexity exists on Arbitrum \cite{Arbitrum}, where block proposals are controlled by a single proposer. Arbitrum's Timeboost \cite{TimeBoostDocs} allows this Arbitrum proposer to outsource block-building through an auction. The winners of this auction are announced ahead-of-time, with winners given an exclusive fixed-time advantage to propose partial blocks before the whole blocks are completed (transactions can be appended to the partial blocks) and proposed by the proposer. Auction winners can use this time advantage to preconf transactions, for themselves or for others at some cost. As such, auction bids reflect the total preconf revenue that can be captured by leveraging this exclusive time advantage. Unfortunately, the algorithms being used to participate meaningfully in Timeboost and its auctions remain closed-source. 
    
    According to \cite{TimeboostAuctionWinnerBreakdown}, more than 90\% of Timeboost auctions are being won by two entities -- an early sign that preconf pricing will follow similar, if not more extreme, centralization trends as those observed in the MEV-Boost markets. For comparison and as previously mentioned, at any given time in the last two years, two to three builders have been responsible for 90\% of blocks proposed through MEV-Boost\cite{MEVBoostShares}.

    \subsection{Summary of Preconf Economics}

    Preliminary research suggests that while preconfs can create new revenue streams for block proposers, especially via inclusion preconfs, there are also scenarios where preconfs complicate or even reduce revenue (particularly for execution preconfs unless managed optimally). Pricing mechanisms for these services are still an open research problem, and practical implementations like Arbitrum’s Timeboost are providing initial data points. The next section discusses various risks that these protocols entail, complementing these economic analyses with broader considerations.

    \section{Preconf Risks} \label{sec:risk}
    This section focuses on the risks, summarized in Table~\ref{tab:preconf_risks}, that accompany the addition of preconf protocols to a blockchain's block-building pipeline.
    The added sophistication and infrastructure that preconfs require compared to existing block-building pipelines bring risks and considerations for all entities. In this section, we detail some of the key risks and considerations that preconfs bring. The headline risks are summarized in Table \ref{tab:preconf_risks}.

    Although intended to be comprehensive, more risks may surface as more preconf protocols emerge and underlying blockchains continue to evolve (e.g., through the introduction of new EIPs for Ethereum~\cite{W:Future-ProofingPreconfirmations}). 
    
    \begin{table}[h!] 
        \centering
        \resizebox{0.8\textwidth}{!}{%
        \begin{tabular}{|>{\raggedright}p{3cm}|c|c|c|c|}
        \hline
             & Preconfer & Proposer & User & Blockchain 
             \\ \hline
                Implementation & \checkmark & \checkmark & \checkmark & \checkmark \\ \hline
                Slashing & \checkmark & \checkmark & & \checkmark \\ \hline
                Reputation & \checkmark & \checkmark & & \checkmark \\ \hline
                Liveness & \checkmark & \checkmark & \checkmark & \checkmark \\ \hline
                Legal & \checkmark & \checkmark & & \\ \hline
                Centralization & \checkmark & \checkmark & \checkmark & \checkmark \\ \hline    
                Congestion & \checkmark & \checkmark & \checkmark & \checkmark \\ \hline         
        \end{tabular}%
        }
        \caption{Preconf Risks}
        \label{tab:preconf_risks} 
    \end{table}

\subsection{Implementation Risk}
    Preconf protocols are additional technological components in an ever-growing blockchain tech stack. Like all blockchain components, preconf protocols must be thoroughly tested and audited to ensure protocols act as intended, are free of bugs, and are resilient to failures of other components in a blockchain's tech stack. Implementation risk increases the likelihood and threat of all other risks, even in scenarios where stakeholders are following the rules of the preconf protocol. The main defense against this is robust protocol design and thorough auditing of preconf protocols.

\subsection{Slashing}
    As described in \hyperref[step1:preconfer_registration]{Section~\ref{step1:preconfer_registration}} and \hyperref[punishments]{Section~\ref{punishments}}, \emph{slashing} is an ex-post, direct penalty applied to slashable collateral locked by preconfers and imposed via enforcement mechanisms (see \hyperref[sec:enforcement]{Section~\ref{sec:enforcement}}). 
    While slashing is meant to disincentivize undesired preconfer behavior in a preconf protocol, there is the potential for unfair or accidental slashing.
    The risk of being slashed is relevant for several entities in a preconf protocol: Preconfers, who post slashable collateral; proposers, who inherit penalties from preconfers; and the underlying blockchain, whose economic security may be affected by mass slashing. 

    \subsubsection{Slashing: Preconfer and Proposer}
    \label{risk_slashing:preconfer}
        The risk of being slashed depends on the type of fault the preconfer or proposer (perhaps unintentionally) committed (see \cite{W:Basedpreconfirmations}): (i) \textbf{Safety fault:} Safety faults should never be committed by an honest preconfer who is also a proposer (see \cite{W:Basedpreconfirmations} and Section~\ref{preconfer_faults_and_punishing_conditions}). 
        When the preconfer is not a proposer, both the preconfer and the proposer can be responsible for a safety fault. The exact way depends on the implementation but an example is the following:
        the preconfer fails to make the proposer aware of any preconfed transactions in time for proposal or the proposer omits specific preconfed transactions at proposal time. To reduce the risk that a party is penalized without performing a fault, a fault attribution mechanism can be implemented. One example is to have a relay honestly report the list of preconfed transactions that was made known to the proposer in time \cite{W:FaultAttribution}; (ii) \textbf{Idleness fault:} If idleness faults are penalized through slashing, a preconfer can be slashed even without malicious intent, for example, if their server is down or there is a network issue and they fail to publish preconfs on time. In a protocol with a permissioned overseer, there is the additional risk that a malicious overseer may unfairly slash the preconfer even when no fault is actually committed; (iii) \textbf{Liveness fault:} For liveness faults, the proposer is a stronger candidate for slashing, as it is more likely to have misbehaved. However, a fault attribution mechanism could again reduce the risk of unfairly penalizing the proposer if the preconfer delayed sending preconfs. 
                The proposer risks being slashed even for missing slots accidentally. To protect themselves against this risk, proposers could request that preconfers chain their preconfs \cite{W:AvoidingAccidentalLivenessFaultsforBasedPreconfs}, allowing preconfs to be fulfilled by subsequent proposers in the lookahead in return for a proportion of the tip revenue. Preconf protocols can amplify the economic impact of missing slots in L1, where proposers already incur the opportunity cost of missing out on proposer rewards and revenue.

    \subsubsection{Slashing: Blockchain}
    \label{risk_slashing:protocol}
        As discussed in Section~\ref{step1:preconfer_registration}, prospective preconfers are required to post slashable collateral. By imposing barriers to entry, preconf protocols:
        \begin{itemize}
            \item Require careful cryptoeconomic protocol design to ensure that incentives are aligned in a way that incentivizes entities to participate.
            \item Risk centralizing the service's provision to a few, well-capitalized entities. 
        \end{itemize}
        The latter has risks for the underlying blockchain especially when slashable collateral is sourced via the capital-efficient means of restaking.
        In this case, a single malicious or accidental act by a preconfer could trigger cascading slashings across any protocols also secured by the restaked collateral. 
        If a widespread slashing event were to occur, the security of both the underlying blockchain and the specific preconf protocol would be affected (e.g., see \cite{risk_restaking} and \cite{W:Restaking101:APrimeronEigenLayer}).
        Therefore, dedicated collateral used by preconfers to register is less risky for the preconf protocol than restaked collateral.
        

\subsection{Reputation} \label{ReputationRisk}
    Unlike the direct financial penalties of slashing, \emph{reputation loss} is an indirect and ex-post punishment (see \hyperref[punishments]{Section \ref{punishments}}) meant to reduce the future revenue of the affected preconfer. 
    As an attribute, an entity's reputation can affect their standing both inside the preconf protocol, which may lead to revenue loss, and outside of the context of a preconf protocol.
    The risk of diminished reputation concerns multiple entities in a preconf protocol: Preconfers, who face direct reputational damage and indirect revenue loss from their own misbehavior; proposers, who may be held accountable for the behavior of preconfers; and the blockchain protocol, which may suffer reputational damage if preconf protocols are perceived by users to be unfair or extractive (e.g., proposers allocate most of the block-space to preconfed transactions, leaving little room for users who do not want to preconf their transaction).
    \subsubsection{Reputation: Preconfer and Proposer}
    \label{risk_reputation:preconfer}
    The reputation of both a preconfer and a proposer can be harmed by any faults they commit. This risk of diminished reputation also arises when a party is perceived as responsible for non-fulfillment even though the other party committed the fault. 
    Even if the party penalized for a safety or liveness fault is the one that actually committed it, users may not take the time to determine which party is truly at fault -- especially in the absence of a fault-attribution mechanism. Diminished reputation could cause users to send fewer future preconf requests to the misbehaving preconfer 
    which would reduce preconf tip revenue. Moreover, if a proposer runs multiple validators, reputation loss can affect all of them. 
    Reputational damage can extend beyond any foregone revenue within the preconf protocol itself to other protocols or business activities in which the preconfer or the proposer is engaged. To mitigate this risk, some preconfers or proposers may choose to operate their services using separate entities. 
    
    \subsubsection{Reputation: Blockchain}
    \label{risk_reputation:blockchain}
        The behavior of entities in and specific implementation of preconf protocols could affect the underlying blockchain's reputation.
        If preconfs are perceived to be highly extractive, for example through DSSAs (see Section~\ref{sec:execpreconfproposerrevenue}), it could negatively affect the user experience and consequently the reputation of the blockchain. 
        In the long run, the loss in the blockchain's reputation could result in (i) an exodus of users and applications in search of a blockchain with better user experience and (ii) a drop in the native currency’s exchange rate.

\subsection{Liveness} 
    Liveness risks affect multiple entities in a preconf protocol: 
    Preconfers, who may be unable to fulfill their preconfs due to proposer downtime; proposers, who risk being slashed and losing rewards for missed slots; and users, whose preconf requests may go unfulfilled.  

    \subsubsection{Liveness: Preconfer}
    A preconfer's ability to fulfill their preconfs depends on the liveness of both the proposer and the underlying blockchain. 
    As mentioned in \hyperref[risk_slashing:preconfer]{Section \ref{risk_slashing:preconfer}}, preconfers can be punished for idleness faults, regardless of whether the preconfer was at fault or not. 

    \subsubsection{Liveness: Proposer}
    \label{risk_liveness:proposer}
    Proposers are responsible for proposing valid blocks. 
    In the context of preconf protocols, a missed slot is punished via slashing and foregone revenue for the proposer (see Section~\ref{risk_slashing:preconfer}).
    Missed slots may happen for reasons both internal and external to the proposer: 
        \begin{itemize}
            \item \textbf{Internal:} As discussed in \hyperref[preconfer_faults_and_punishing_conditions]{Section~\ref{preconfer_faults_and_punishing_conditions}}, the proposer may go offline for any number of (potentially accidental) reasons -- such as power outages or internet downtime -- making them unable to propose. Solutions to allow proposers to insure themselves against accidental liveness faults through chaining have been discussed in the same section. 
            \item \textbf{External:} If the proposer, or their delegated gateway, outsources final block construction (see Section~\ref{delivery_tradeoffs}), the proposer risks the full block containing preconfed transactions not being propagated in time. 
        \end{itemize}
    From the perspective of the preconf protocol, these cases are indistinguishable. 
    As with the fair attribution of safety faults described in Sections~\ref{risk_slashing:preconfer} and~\ref{risk_slashing:preconfer}, in the case of liveness faults caused by external factors, exonerating the proposer will likely require the involvement of a trusted third party. 
    
    \subsubsection{Liveness: User} 
    Users specify a preconf tip to the preconfer in expectation of preconf fulfillment (see Section~\ref{preconf_request}).
    Preconf fulfillment mainly depends on the preconfer and proposer being online and non-malicious:
        \begin{itemize}
            \item \textbf{Preconfer:} If the preconfer is found by an overseer to be committing an idleness fault (see Section~\ref{preconfer_faults_and_punishing_conditions}), users' preconf requests will remain unfulfilled, and, if the preconfer does not come back online, may eventually expire. 
            \item \textbf{Proposer:} As described in Section~\ref{risk_liveness:proposer}, the proposer may commit a (potentially accidental) liveness fault for both internal and external reasons. If the users' preconfs were chained, they will be fulfilled, only in a later slot. If not chained, users' preconfs will remain unfulfilled and will eventually expire.
        \end{itemize}
    The cryptoeconomic security of the preconf protocol has a direct impact on the risks faced by users. For instance, users are better protected when the financial penalties imposed on the preconfer and proposer for failing to fulfill the preconf exceed any potential rewards from misbehavior. Additionally, the risk that users face is closely related to the costs incurred by a malicious preconfer or proposer whose sole intent is to disrupt the preconf protocol. When penalties are inadequate, the consequences can be significant \cite{W:MEVLawsuit, W:TotalEclipseOfTheRelay}. Finally, the risk for users can be mitigated if users are compensated for non-fulfillment, e.g., in \cite{W:Documentation-Understandingmev-commit}. 
    
    \subsubsection{Liveness: Blockchain} 
\label{risk_liveness:protocol}
In some cases, liveness faults in preconf protocols may disrupt the liveness of the underlying blockchain, for example: 
\begin{itemize}
    \item When a proposer delegates preconfing to a gateway, and the gateway delays -- or fails entirely -- to deliver the preconfs, the proposer may be delayed in proposing their block, potentially resulting in the block being excluded from the blockchain.
    \item When the preconfer is also the proposer, then:
    \begin{itemize}
        \item Their downtime can affect both the liveness of the preconf protocol and the liveness of the underlying blockchain.
        \item If the cryptoeconomic design of the preconf protocol creates incentives for a preconfer -- who also acts as a proposer -- to intentionally miss a slot (e.g., when they are scheduled to propose two consecutive blocks and can extract more MEV by skipping the first), it can negatively impact the liveness of the underlying blockchain.
    \end{itemize}
\end{itemize}

\subsection{Legal Risk}
Preconfs are contracts between the user and the preconfer. 
As such, contract law \cite{W:ContractLaw} may apply when arbitrating failed fulfillment of a preconf.
This presents a potential legal risk for proposers and preconfers with respect to preconfing. The exact legal implications of preconfs will emerge as preconf protocols mature.

\subsubsection{Legal: Preconfer} 
    A failure to fulfill preconfs could find users seeking recompense for damages caused by the preconfer.
    The legal entities behind preconfers may be held liable through legal channels if a failure to fulfill a preconf results in financial harm to a user. Recall that a preconf is a commitment that the blockchain ledger at a future time satisfies some predicate. 
    Users, as a result of receiving a preconf, may decide to initiate other actions.
    For example, a preconf to buy ETH on a decentralized exchange may cause a user to sell ETH on a centralized exchange in an attempt to profit from arbitrage. If the preconf does not get fulfilled, users may be unable to retroactively cancel their sell order, potentially resulting in financial loss.  
    
    To mitigate the risk of preconfers being targeted for recompense through legal systems, preconfers may choose to include explicit disclaimers about lack of liability for any losses incurred by users as a direct or indirect result of interacting with the preconf protocol.

    \subsubsection{Legal: Proposer}
    As described in Sections~\ref{risk_slashing:preconfer} and~\ref{risk_reputation:preconfer}, proposers that delegate preconf duties inherit some of the risks that preconfers face. If a delegated preconfer fails to fulfill its preconfs, the proposer could likewise face legal challenges due to user damages.

\subsection{Centralization} 
\label{risk_centralization}
Preconf protocols exert centralizing pressures on blockchains. This can be seen through the simultaneous increase in revenue that preconfs offer to preconfers (see \hyperref[preconf_economics]{Section \ref{preconf_economics}}) and the increased sophistication required to preconf vs not preconfing (see \hyperref[preconfs_impact_proposer_revenue]{Section \ref{preconfs_impact_proposer_revenue}}. 
This combination incentivizes blockchain entities to offer preconfs, which either means outsourcing preconfing (and at least partial block-building) to professional entities with economies of scale, or the blockchain entities themselves becoming preconfers, managing all necessary preconf infrastructure, pricing, and risk independently. Regardless of which of these paths emerge, preconfs risk the unexpected and likely undesired consolidation of blockchain roles into a small set of highly-sophisticated entities. 

In the following sub-sections we explore in more detail the centralization pressures, effects, and mitigations being considered for each of the blockchain entities we consider.

    \subsubsection{Centralization: Preconfer}

    The centralization risk to preconfers from preconfing centers around the expected competitiveness of preconfing. Preconfing's sophistication requirements likely means consolidation of the preconfer set into a small number of dominant players. This means tight profit margins and high barriers to entry for all but the most sophisticated preconfers.

    \subsubsection{Centralization: Proposer}
    
    The potential revenue from preconfs could force proposers to participate in preconf protocols. For low-resource/unsophisticated proposers, this likely means delegation of preconfing to a small group of gateways/builders in the same way that block-building is outsourced today. Given the added complexity of preconfing compared to normal block-building, this small group of delegates may become dominated by one or two oligopolies, in-line with the centralization seen in MEV-Boost\cite{MEVBoostShares}. Oligopolistic delegates can elect to extract rents from proposers, especially if preconfing becomes ingrained in a blockchain's block-building pipeline. These oligopolies also become points of failure, which could prevent a proposer from proposing if delegate failure were to occur. 

    One proposed direction towards mitigating the risks of oligopolies forming among preconfer delegates is the preconfirmation sauna \cite{W:ThePreconfirmationSauna}. In this proposal, the authors advocate for the development of generalized preconfirmation infrastructure and tooling to enable preconfers and proposers to reduce the friction of interactions and on-boarding of entities to preconf protocols. The prevents vendor lock-in, which is a key factor in enabling oligopoly abuses.

    \subsubsection{Centralization: User}

    As the blockchain ``consumer'', users face a degraded user experience if centralization through preconfs materialize. Users risk degradation across each of censorship resistance, fee increases, and fault tolerance. These effects are all synonymous with service provider monopolization.

    Apart from centralization of the block-building pipeline being a risk for users, there are also centralization pressures on users themselves. Users must calculate appropriate preconf tips, customs penalties (see \cite{W:User-DefinedPenalties:EnsuringHonestPreconfBehavior}), create valid preconf requests, and identify the correct preconfers or endpoints to send their requests to.
    
    In the case of preconf tips, Section \ref{preconf_economics} described how tips are expected to be more complex and volatile than regular EIP-1559 transactions. This makes the setting of preconf tips potentially dangerous, which in turn means likely outsourcing and additional fees for these services.

    \subsubsection{Centralization: Blockchain}

    If the underlying blockchain does not expect a blockchain entity to engage in preconfs, there is likely some centralization being introduced that the blockchain is not designed for. 

    While centralization may be an acceptable design trade-off for scalability in many L2s (see Section~\ref{sec:intro_L2}), blockchains that are designed to have decentralized and permissionless roles may see centralization as unacceptable. This is particularly true for Ethereum. Without proposer decentralization, Ethereum's censorship resistance and fault tolerance guarantees fail \cite{W:DecentralizedCryptoNeedsYou}. Unfortunately, these guarantees are already failing, as even now, proposers are outsourcing block-production en-masse to the same two or three entities through MEV-Boost \cite{MEVBoostShares} (see Section \ref{sec:L1_pipeline}). Preconfs stand to exacerbate this risk for decentralized blockchains, given the additional complexity that preconfs entail (see Section \ref{sec:price}). 

    The protection closest to being deployed  against centralization effects in Ethereum is FOCIL (fork-choice enforced inclusion lists), with a full specification available under the heading of EIP-7805 \cite{EIP7805}, and incentive analysis of candidate transaction fee mechanisms in \cite{stouka2025multipleproposertransactionfee}. FOCIL allows non-proposing validators to enforce transaction lists upon block proposals in Ethereum, assuming that non-proposing validators will remain decentralized and majority honest. This is despite the evidenced rationality of validators when elected as proposers, through their active use of MEV-Boost. As such, FOCIL alone stands as a temporary fix for an ever-centralizing Etheruem validator set. 
    
    To tackle the misalignment of proposer incentives with other validator incentives, several protections have been discussed which aim to separate the role of the proposer from other validator roles, most notably attesting, but potentially also participation in FOCIL-like protocols for censorship resistance \cite{W:UnbundlingStaking:TowardsRainbowStaking, W:Three-TierStaking, W:AppointedExecutionProposers}. Worryingly, interest in these solutions has been small, with development pending, to the best of our knowledge. Given the threat of centralization to decentralized systems is large, this demands increased awareness, investigation, analysis and development of centralization protections.

\subsection{Congestion}
    Preconf protocols put the underlying blockchain and blockchain infrastructure at risk of becoming congested. This is for two interrelated reasons: 
        \begin{itemize}
            \item Entities are incentivized to vertically integrate with preconfers to maximize their ability to act on contentious state \cite{W:StrawmanningBasedPreconfirmations}.
            \item Users may spam preconfers with requests to address the randomness of delivery times and value of interacting with contentious state first \cite{W:StrawmanningBasedPreconfirmations,W:BasedPreconfirmationswithMulti-roundMEV-Boost}.
        \end{itemize}
        
    This heavy traffic strains infrastructure, increases costs, and in the case of spam, results in wasted blockchain resources. 
    The existing block-building pipeline isolates and contains congestion risk on the builder and relay side, with the proposer only required to select block-headers, and the resulting blocks containing little-to-no failed transactions in order to maximize a builder's revenue (builders do not receive fees from failed transactions).
    
    Congestion impacts all preconf protocol entities: Proposers and preconfers, who are required to understand how congestion impacts revenue and how to handle worst-case loads; users, whose costs and inclusion/execution time may become more volatile; and the underlying blockchain and its applications, which may not be equipped to handle large amounts of traffic. In addition to creating inequalities among users, latency races can incentivize geographical centralisation~\cite{W:StrawmanningBasedPreconfirmations}.  A potential mitigation to congestion risk is an auction-based system that batches preconfs in sub-slots instead of streaming continuously~\cite{W:BasedPreconfirmationswithMulti-roundMEV-Boost, W:AnalysingExpectedProposerRevenuefromPreconfirmations}. Through batching, proposers and preconfers can outsource congestion management to builders, a technique already employed in Ethereum's block-building pipeline (see Section \ref{sec:L1_pipeline}).

\section{Existing Protocol Case Studies}
\label{sec:implementations}
    At this point of the SoK, we are equipped with sufficient generalized preconf knowledge to examine several in-production preconf protocols in detail. For ease of reading and consistency, we map each of the examined protocols to the six-step preconf pipeline framework from \hyperref[sec:pipeline]{Section \ref{sec:pipeline}}.

    \subsection{ \textbf{Optimism}}
        This subsection examines the preconfs inherent in a non-based L2 with a centralised proposer (see \hyperref[sec:intro_L2]{Section \ref{sec:intro_L2}}), using Optimism~\cite{Optimism,W:OptimismDocs} as a case study. Although not explicitly labeled a ``preconf protocol'', the confirmations provided by the proposer serve the same practical function for users, offering guarantees of the blockchain ledger before it is posted to L1.
            
        \begin{enumerate}
            \item \textbf{Registration:}
            The registration process is permissioned and centralized. There is a single entity named Optimism's sequencer which is operated by the Optimism Foundation. There is no explicit preconfer registry or collateral requirement. 
            
            \item \textbf{Election:}             
            The election is trivial and predetermined. As a network with a single, designated block producer, the sequencer is implicitly the elected preconfer for every L2 block.
            
            \item \textbf{Request:}
            Users who wish to receive a sequencer confirmation submit a standard signed L2 transaction directly to the sequencer's private mempool.
            
            \item \textbf{Response:}
            Upon receiving a transaction, the sequencer determines the transaction's execution order, computes its resulting state change, and includes it in an ``unsafe'' L2 block. 
            The response to the user is a high-fidelity execution preconf. It is not merely a promise of future inclusion but a firm commitment to a specific ordering and outcome, complete with a post-transaction state root (a commitment to the L2 blockchain ledger as it will be formed after the inclusion of these transactions in this specific order). Nevertheless, unsafe transactions may not be finalized if the sequencer fails to publish the block to Ethereum within the sequencing window or if it re-organises the unsafe blocks.
            
            \item \textbf{Fulfillment:}
            Fulfillment occurs in two stages. First, a separate batcher process compresses transaction data from unsafe blocks and submits it to the L1, at which point the L2 block status is upgraded from ``unsafe'' to ``safe''. Second, once the L1 block containing this data is finalized by Ethereum's consensus (meaning that it is included in the Ethereum blockchain ledger), the L2 transaction is considered ``finalized'', completing the sequencer's promise.
            
            \item \textbf{Punishment:} 
            The punishment mechanism is primarily indirect and ex-post (see \hyperref[punishments]{Section \ref{punishments}}). There is no automated on-chain process to slash the centralized sequencer for misbehavior. Instead, sequencer punishment relies on reputational damage and the threat of users migrating away from the ecosystem. This model is a stepping stone, as the long-term roadmap for the OP Stack ecosystem includes plans for a decentralized set of sequencers.
        
        \end{enumerate}

        \textbf{Fair-exchange} (see \hyperref[fair_exchange_problem]{Section \ref{fair_exchange_problem}}) of preconfs is trusted to take place. 

     \subsection{\textbf{Taiko's ``Permissionless Preconfs''}}
        This subsection introduces the design of \emph{permissionless preconfs} for Taiko, a based rollup, as a concrete case study of how based rollups can implement based preconf mechanisms\footnote{Public documentation, to be posted here: \url{https://github.com/taikoxyz}, missing at time of writing. This will be updated as soon as the documentation is made publicly available.}.
        
        \begin{enumerate}
        \item \textbf{Registration:} Happens through the Universal Registry Contract (URC) (see \hyperref[preconfer_registry]{Section \ref{preconfer_registry}}).
        \item \textbf{Election:} Uses an \emph{optimistic lookahead scheme}: the first preconfer of each epoch posts the lookahead for the next epoch, consisting of those in the L1 proposer lookahead who have opted in to Taiko preconfs, which anyone can challenge with a fraud proof if incorrect. EIP-7917 \cite{EIP7917} is used during these fraud proofs to verify that the submitted lookahead matches the canonical proposer schedule (see \hyperref[sec:L1_pipeline]{Section \ref{sec:L1_pipeline}}). If no proposer is registered, a fallback preconfer is used.
        \item \textbf{Request:} Standard L2 transactions submitted through the public L2 mempool.
        \item \textbf{Response:} The elected preconfer collects transactions from the public L2 mempool, builds an L2 block, and publishes a signed commitment to the hash of the transaction list together with the transaction list itself every two seconds. This interval is a protocol parameter and may be adjusted over time. Because this commitment specifies the full set of transactions in their canonical execution order, L2 client software can locally execute them to derive the latest L2 state -- making this effectively an execution preconf (see \hyperref[def:execution_preconf]{Definition \ref{def:execution_preconf}} and \hyperref [subsec:transaction_based]{Section \ref{subsec:transaction_based}}). To explicitly end their preconf duties for the slot, the preconfer issues an \emph{end-of-preconf} message, which commits to stop preconfing and hands over preconf duty to the next preconfer.
        \item \textbf{Fulfillment:} The preconfer must include the preconfed transaction list in their transaction batch posted to L1 before their elected window closes, either in their own L1 block or via the L1 public mempool.
        \item \textbf{Punishment:} Triggered when there is a mismatch between what was sent in the response and what was eventually fulfilled. Any entity can slash a preconfer for mismatched transaction lists, missed submissions, or equivocating on end-of-preconf message (e.g., publishing an additional L2 block after issuing an end-of-preconf message).
        \end{enumerate}
        
        To address the \textbf{fair-exchange problem}, Taiko employs an overseer, governed by the Taiko DAO, that monitors preconf timing and can blacklist operators who delay publication, incentivizing timely preconfs.    

    \subsection{\textbf{Primev's mev-commit Protocol}}
    \label{subsec:primev}
    In this section, we describe how the mev-commit protocol~\cite{W:MevCommitWhitepaper,W:Documentation-Understandingmev-commit} issues preconfirmations, or \emph{commitments} in the preconf protocol's terminology. While we follow the general structure introduced in Section~\ref{sec:pipeline}, it is important to note that sometimes mev-commit slightly departs from that framework. 

    The central idea of mev-commit is to enable L1 block builders to provide users with enforceable, yet conditional, assurances about transaction inclusion or execution before the corresponding L1 block is built. To achieve this, the protocol maintains a dedicated mev-commit blockchain with fast block times. When a user requests a preconf by submitting a bid, a block builder may issue a conditional preconf where the condition is that the builder is selected to build the specified L1 block. 
    
    To issue such a preconf, the builder records a cryptographic commitment on the mev-commit chain. These commitments hide all details of the bid but can later be opened to allow public verification and settlement. The commitments are supposed to be opened by the block builder after building the corresponding L1 block. If the builder fails to open, e.g., because the promised conditions were not met, the user can instead open the commitment to prove non-fulfillment. In this way, mev-commit enforces accountability for preconfs without sacrificing confidentiality at the bidding stage.

\begin{enumerate}
    \item \textbf{Registration:} Block builders enroll in the protocol through the \emph{Provider Registry}, a smart contract deployed on the mev-commit chain.

    \item \textbf{Election:} The protocol does not implement an election mechanism. Instead, providers issue conditional preconfs that are valid only if the block they build is annexed to the L1 chain at the specified height.

    \item \textbf{Request:} Users may request different types of conditional preconfs, ranging from simple inclusion preconfs to execution preconfs.

    \item \textbf{Response:} Upon receiving a preconf request, a block builder may ignore it or issue a preconf by publishing a cryptographic commitment to the mev-comit chain. The request specifies all conditions for fulfillment. Importantly, the associated bid (the fee a user offers for a preconf) decays over time from submission, incentivizing builders to respond quickly to maximize profit.

    \item \textbf{Fulfillment:} When a builder $B$ produces an L1 block at height $h$, all preconfs issued by $B$ for height $h$ become enforceable. If $B$ fails to honor any such preconfs, penalties apply.

    \item \textbf{Punishment:} The penalty for failing to fulfill a preconf is specified in the request as a compensation amount the user will receive in such cases. If the preconf is not honored, this compensation is transferred to the user, and the preconf bid is refunded.
\end{enumerate}

     To address the \textbf{fair-exchange problem}, the protocol utilises a tip decay mechanism, and an oracle that monitors the L1 chain. The actual tip amount is computed using the timestamp of when the commitment was issued as recorded on the mev-commit chain. This incentivizes builders to issue preconfs in a timely manner after receiving the bid. The oracle then looks at opened commitments and checks whether the conditions for fulfillment are met. If so, the builder receives the preconf tip specified in the bid. If not, the builder is penalized by transferring the compensation amount specified in the bid to the user

        \subsection{ \textbf{ETHGas}}
        ETHGas ~\cite{W:IntroducingETHGasandRealtimeProposerCommitmentstotheLidoCommunity,W:ETHGasDocs-Overview} is a preconf protocol which allows upcoming Ethereum proposers to sell commitments to their block-space. ETHGas is integrated with existing PBS infrastructure (see \hyperref[proposersAndAttesters]{Section \ref{proposersAndAttesters}}) using Commit-Boost~\cite{W:ETHGaspreconfcommit-boostmodule}, a modified version of the standard MEV-Boost~\cite{MEV-Boost} client that enables ETHGas stakeholders to create and manage preconfs.

        \begin{enumerate}
            \item \textbf{Registration:}
            L1 proposers wishing to register with the ETHGas protocol must run Commit-Boost and also post collateral. The collateral can be posted by restaking using EigenLayer (see \hyperref[preconfer_registry]{Section \ref{preconfer_registry}}) or deposited directly into the ETHGas collateral smart contract~\cite{W:ETHGasDocs-Validators}.
            
            \item \textbf{Election:} 
            ETHGas does not run its own election process but rather provides a platform for these pre-determined proposers to sell commitments for the block-space they are already entitled to produce. Furthermore, a preconfer can choose to sell inclusion preconfs, execution preconfs or whole block commitments~\cite{W:IntroducingETHGasandRealtimeProposerCommitmentstotheLidoCommunity,W:ETHGasDocs-Overview}.
            
            \item \textbf{Request:}
            Users submit requests and purchase preconfs through the API of ETHGas Exchange.  Proposers can sell preconfs directly through ETHGas Exchange while less sophisticated proposers may choose to delegate preconfing process to a third party. Inclusion preconfs can be purchased up to 32 slots in advance, while whole block commitments can be secured up to 64 slots in advance~\cite{W:ETHGasDocs-CommitmentBuyers,W:IntroducingETHGasandRealtimeProposerCommitmentstotheLidoCommunity}.
                        
            \item \textbf{Response:}
            After a request is received and accepted, the elected L1 proposer for the target slot uses their validator key to sign a cryptographic commitment to the user's request. This signed message, which is delivered to the buyer via the ETHGas Exchange, serves as the ~\cite{W:IntroducingETHGasandRealtimeProposerCommitmentstotheLidoCommunity,W:ETHGasDocs-TheETHGasArchitecture}.
            
            \item \textbf{Fulfillment:}
            The ETHGas Exchange streams preconfs to block builders. The builders then construct a valid block that respects these commitments and submit it to a relay. The proposer fetches the header of the most profitable block from the relay, signs it, and submits it to the network. The relay then publishes the full block body. The relay can also deliver a fall-back block in case a builder fails to produce a conforming block in time~\cite{W:ETHGasDocs-BuildersSequencers,W:ETHGasDocs-Relays}.
            
            \item \textbf{Punishment:}  
            After a block is finalized, the ETHGas back-end system automatically verifies if the proposer fulfilled their preconfs. It checks for a verifiable on-chain discrepancy between the signed preconfs and the actual content of the block proposed for that slot (or if the slot was left empty). If a violation is detected, the system initiates a process to slash the proposer's collateral, compensating the affected user~\cite{W:ETHGasDocs-Validators}.
            
        \end{enumerate}

\section{Conclusion} \label{sec:conclusion}

As blockchain systems evolve, preconfs have emerged as a key mechanism for providing users with guarantees during the waiting period between submitting a transaction to the P2P network and its inclusion in the blockchain ledger. By doing so, preconfs bridge the gap between the scalability of the underlying blockchain protocol and the practical needs of end-users and applications. This SoK outlines the fundamental structure of preconf protocols and presents several preconf protocols currently in production. Furthermore, it reviews existing research and highlights areas for further study in the economics of preconf protocols, such as the revenues earned by preconfers and the pricing mechanisms through which users tip them. Finally, it discusses potential risks faced by all entities involved in preconf protocols, as well as implications for the underlying blockchain protocol. \par    
While no single approach to preconfs has become dominant, preconf protocols in general are likely to play a increasingly important role in shaping the evolution and adoption of Ethereum and other blockchain ecosystems, particularly in latency-sensitive use-cases.
Ultimately, the path forward for preconfs requires further research and development. Preconf adoption introduces new complexities, such as more complex economic incentives and trust assumptions, which must be carefully analyzed and balanced. Collaboration across research, preconf protocols, and underlying blockchain systems will be essential to ensure that preconfs enhance user experience without compromising the security of the blockchain protocols. In this regard, preconfs are not merely an optimization, but a foundational element in the broader effort to make blockchains truly usable at scale.

\section*{Acknowledgments}

We thank Christian Matt and Bernardo Magri, both Primev, for their contributions to the SoK. This SoK has received funding from Taiko, and the Ethereum Support Program -- Grant ID FY25-2142. Special thanks to Taiko whose pioneering mission and ongoing successes in implementing preconfirmations inspired this work.

\subsubsection*{Legal Disclaimer}

\textit{This article has been prepared for the general information and understanding of the readers. No representation or warranty, express or implied, is given by Nethermind as to the accuracy or completeness of the information or opinions contained in the above article. No third party should rely on this article in any way, including without limitation as financial, investment, tax, regulatory, legal, or other advice, or interpret this article as any form of recommendation.}

\newpage
\bibliographystyle{IEEEtran}

\bibliography{manual_references}
   \appendix
    \section{Intents and Intent Preconfs}
    \label{Appendix_A}
    In an intent-based architecture, users submit intents instead of fully specified transactions (e.g., see \cite{W:Intent-BasedArchitectureandTheirRisks, ERC-7521}). Intents are artifacts that describe a user's desired outcome without prescribing the exact execution path. More specifically, an intent includes a predicate that defines the conditions under which the user permits their funds to be moved (see \cite{Anoma}). For example, a user might express a willingness to exchange at most $X$ units of token $A$ for $Y$ units of token $B$, without specifying the particular route or counterparty. Entities known as solvers collect these intents and bundle them into a single transaction, which they then submit to the blockchain for inclusion. By aggregating multiple intents -- including potentially their own -- solvers may extract MEV. \par 
    The structure of intents, their execution logic, the nature of solvers, and the incentive mechanisms (such as tips sent by the users or MEV extraction) vary depending on the specific intent-based architecture. While the design of preconfs in these systems remains largely unexplored, to the best of our knowledge, intent requests should resemble preconf requests. Moreover, intent solvers may act as preconfers who commit to fulfilling intents; if they succeed, they receive a reward (e.g., see \cite{cow_intent}), but if they fail, they may be slashed (see \cite{slash_solver}).




\end{document}